\documentclass[prb,twocolumn,aps,superscriptaddress]{revtex4-2}
\usepackage{graphicx}
\usepackage{dcolumn}
\usepackage{bm}
\usepackage{amssymb}
\usepackage{amsmath}
\usepackage{subfigure}
\usepackage{physics}
\usepackage{sublabel}
\usepackage[utf8]{inputenc}

\usepackage{hyperref}
\usepackage{floatrow}
\usepackage{graphicx}
\usepackage{bbm}
\hypersetup{colorlinks = true, linkcolor=blue, citecolor=blue, urlcolor=}
\usepackage[T1]{fontenc}

\begin{document}


\title{Can non-local conductance spectra conclusively signal Majorana zero modes? \\ Insights from von Neumann entropy
}

\author{Abhishek Kejriwal}
\affiliation{Department of Physics, Indian Institute of Technology Bombay, Powai, Mumbai-400076, India}

\author{Bhaskaran Muralidharan}
\affiliation{Department of Electrical Engineering, Indian Institute of Technology Bombay, Powai, Mumbai-400076, India}
\email{bm@ee.iitb.ac.in}

\date{\today}
\begin{abstract}
The topological origin of the zero bias conductance signatures obtained via conductance spectroscopy in topological superconductor hybrid systems is a much contended issue. Recently, non-local conductance signatures that exploit the non-locality of the zero modes in three terminal hybrid setups have been proposed as means to ascertain the definitive presence of Majorana modes. The topological entanglement entropy, which is based on the von Neumann entropy, is yet another way to gauge the non-locality in connection with the bulk-boundary correspondence of a topological phase. We show that while both the entanglement entropy and the non-local conductance exhibit a clear topological phase transition signature for long enough pristine nanowires, non-local conductance fails to signal a topological phase transition for shorter disordered wires. While recent experiments have indeed shown premature gap-closure signatures in the non-local conductance spectra, we believe that the entanglement entropy can indeed signal a genuine transition, regardless of the constituent non-idealities in an experimental situation. Our results thereby point toward furthering the development of experimental techniques beyond conductance measurements to achieve a conclusive detection of Majorana zero modes.
\end{abstract}

\maketitle

Rashba nanowire-superconductor hybrid systems \cite{Alicea-2010,Sau-2013,Stanescu_2013,PhysRevLett.105.077001,PhysRevLett.104.040502,Jiang_2013,Wilczek2012} are preeminent candidates for detecting and manipulating Majorana zero modes (MZMs) \cite{kitaev:physusp2001}, with the ultimate goal of actualizing fault-tolerant topological quantum computation \cite{Sarma2015,Aasen-2016,obrien:prl2018,RevModPhys.80.1083}. Most experimental efforts aimed at the unambiguous detection of MZMs \cite{Das2012,Mourik-2012,Deng-2016,PhysRevLett.110.126406,Scaling_ZBP_Marcus,Albrecht2016,Scaling_ZBP_Marcus} are based on observing quantized zero-bias conductance peaks (ZBCPs) in two terminal normal-topological superconductor (N-TS) links. However, there has been intense scrutiny and debate due to the non-topological origins of the observed ZBCPs \cite{PhysRevB.100.045302,PhysRevLett.109.267002,PhysRevLett.109.227005,PhysRevB.96.201109}. Near-zero energy Andreev bound states (ABSs), often termed as quasi-MZMs, mimic most of the local MZM signatures \cite{10.21468/SciPostPhys.7.5.061,PhysRevB.86.100503,PhysRevB.97.155425,PhysRevB.96.075161,Lobos}, which calls for other methodologies to distinguish between trivial and topological zero-energy modes. Recent proposals have focused on exploiting the non-locality of MZMs via non-local transport measurements on three-terminal normal-topological superconductor-normal (N-TS-N) setups \cite{PhysRevB.96.195418,PhysRevB.88.180507,Akhmerov,Flensberg_Nonlocal,Puglia_Cond_Matrix,puglia}. These non-local signatures could augment the certainty of MZM detection by gauging the non-trivial correlations in the system. Theoretical claims have presented the ability of non-local conductance spectra to detect topological phase transition based on the closing of the bulk gap at the phase transition point \cite{Akhmerov}. However, experiments present contradicting results \cite{puglia} with the gap closing well before the topological phase transition point. The object of this Letter is to  gauge the effectiveness of non-local conductance spectra itself by considering a three terminal setup based on the  archetypical TS, the Kitaev chain, in conjunction with the topological entanglement entropy. \\

\begin{figure}[!htbp]
	\centering
	\subfigure[]{\includegraphics[width=0.9\textwidth]{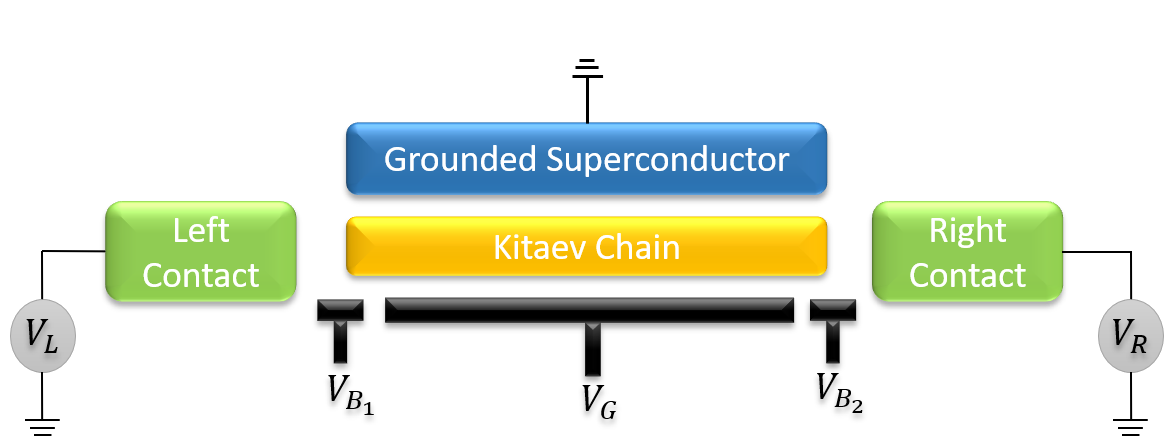}\label{1a}}
	\quad
	\subfigure[]{\includegraphics[width=0.45\textwidth]{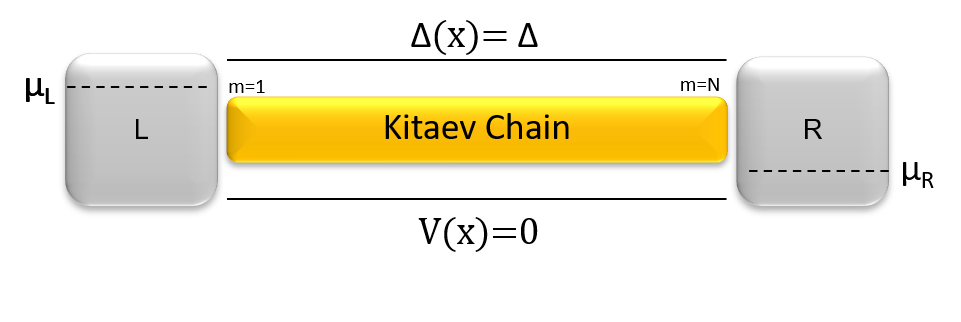}\label{1b}}
	\quad
	\subfigure[]{\includegraphics[width=0.45\textwidth]{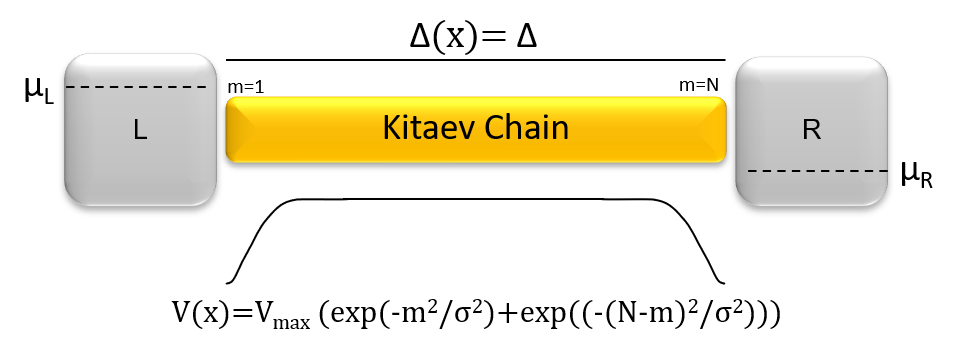}\label{1c}}
	\quad

	\caption{The Experimental Setup. (a) The N-TS-N system considered in this work. The topological superconductor is the archetypical Kitaev chain connected to the ground. The normal contacts are connected to potential $V_L$ and $V_R$ respectively. Three electrostatic gates dictate the potential profile of the setup. Two gate voltages represented by $V_{B1}$ and $V_{B2}$ control the tunneling into the Kitaev chain and the third gate voltage $V_G$ controls the on-site potential $\mu$ in the Kitaev chain. (b) The pristine setup with no disorder in the on-site potential. The disorder potential is zero throughout the chain leading to a homogenous on-site potential. (c) The disordered system with local inhomogeneity in on-site potential. The disorder potential is gaussian and appears at the contacts. Such an inhomogeneity occurs due to Fermi energy mismatch as well as charge inhomogeneities in the system}
	\label{fig:1}
\end{figure}

\begin{figure*}[!htbp]
	\centering
	\subfigure[]{\includegraphics[width=0.22\textwidth]{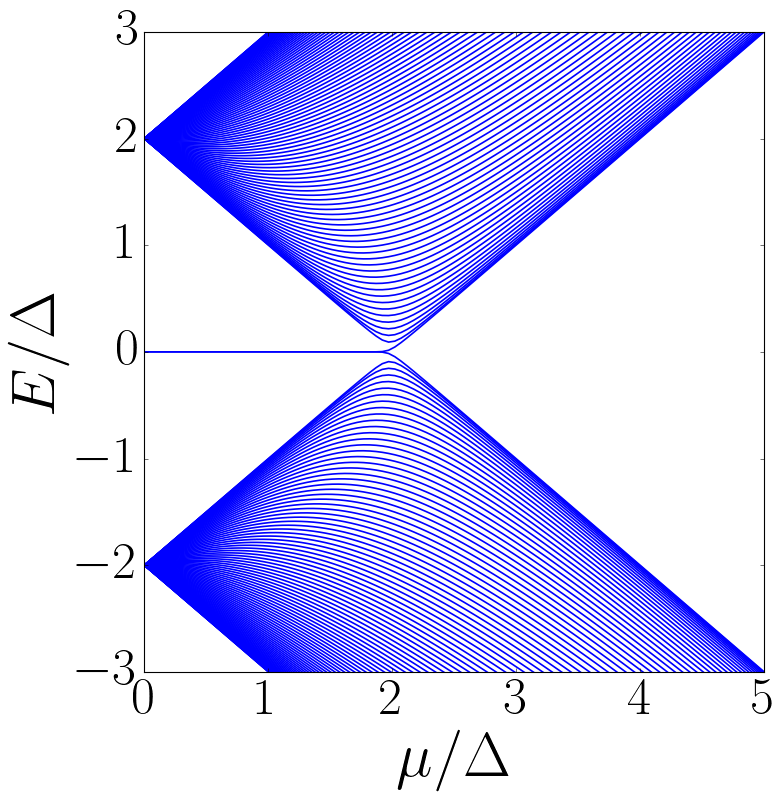}\label{2a}}
	\quad
	\subfigure[]{\includegraphics[width=0.22\textwidth]{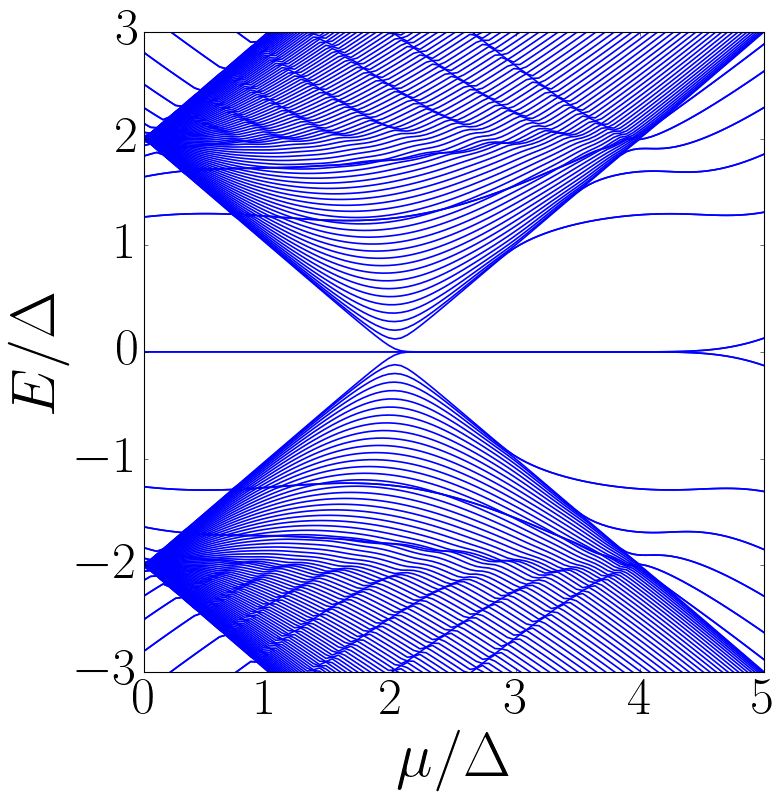}\label{2b}}
	\quad
	\subfigure[]{\includegraphics[width=0.23\textwidth]{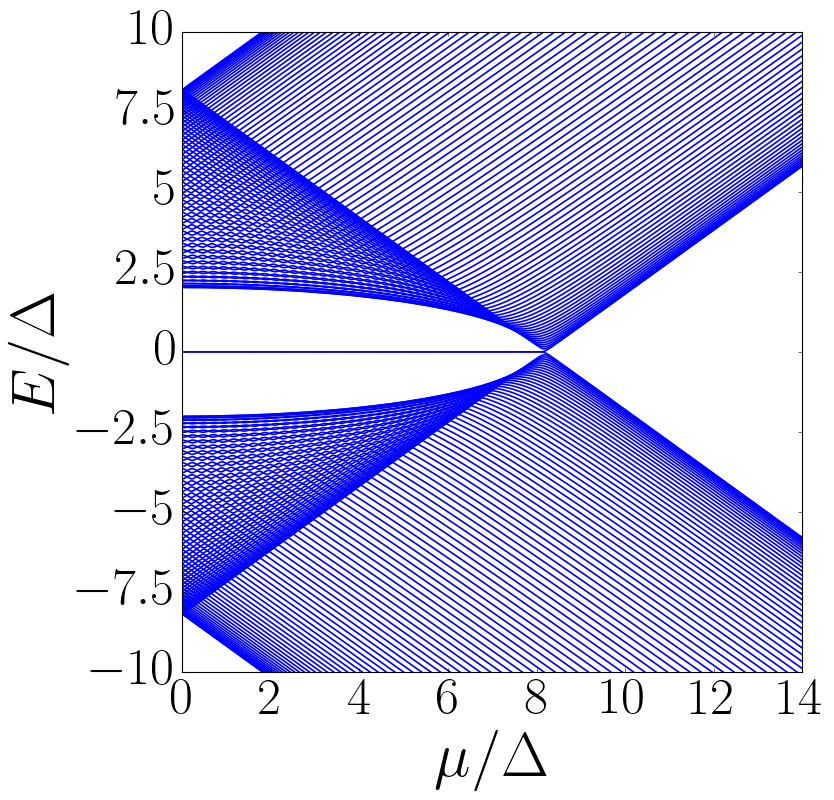}\label{2c}}
	\quad
	\subfigure[]{\includegraphics[width=0.23\textwidth]{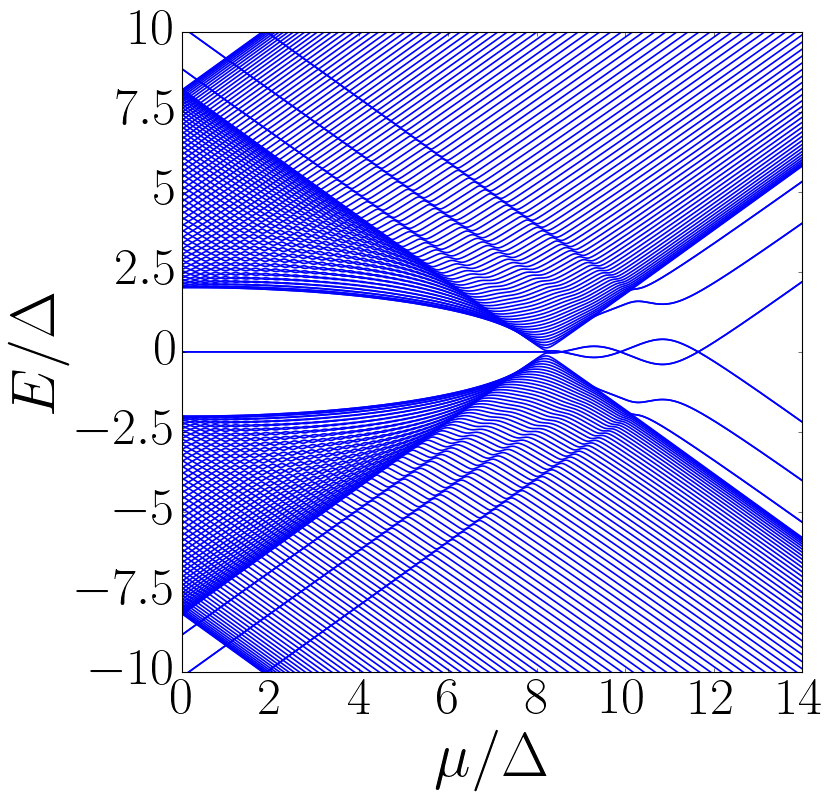}\label{2d}}
	\quad	
	\subfigure[]{\includegraphics[width=0.22\textwidth]{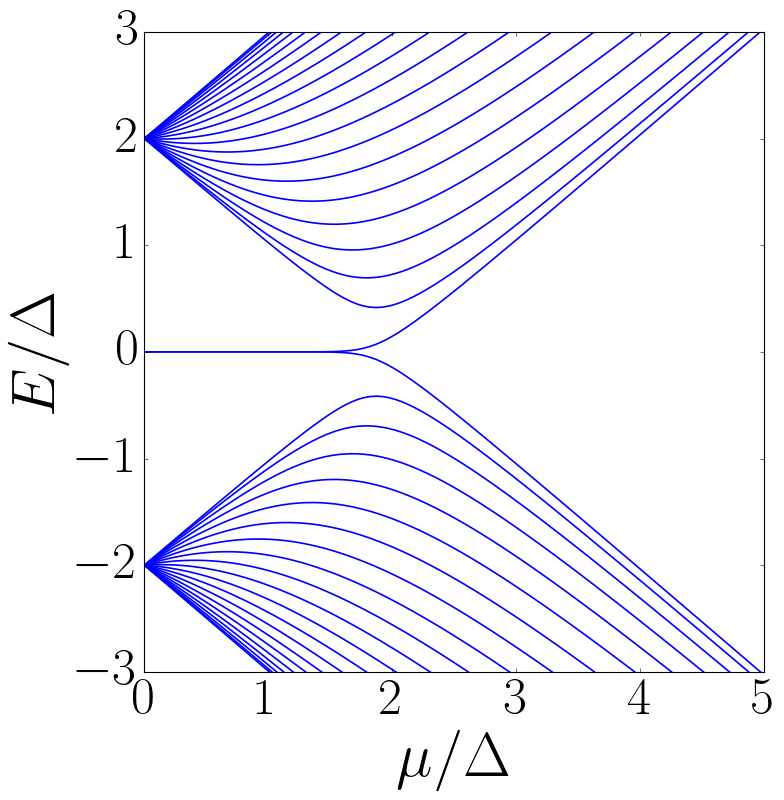}\label{2e}}
	\quad
	\subfigure[]{\includegraphics[width=0.22\textwidth]{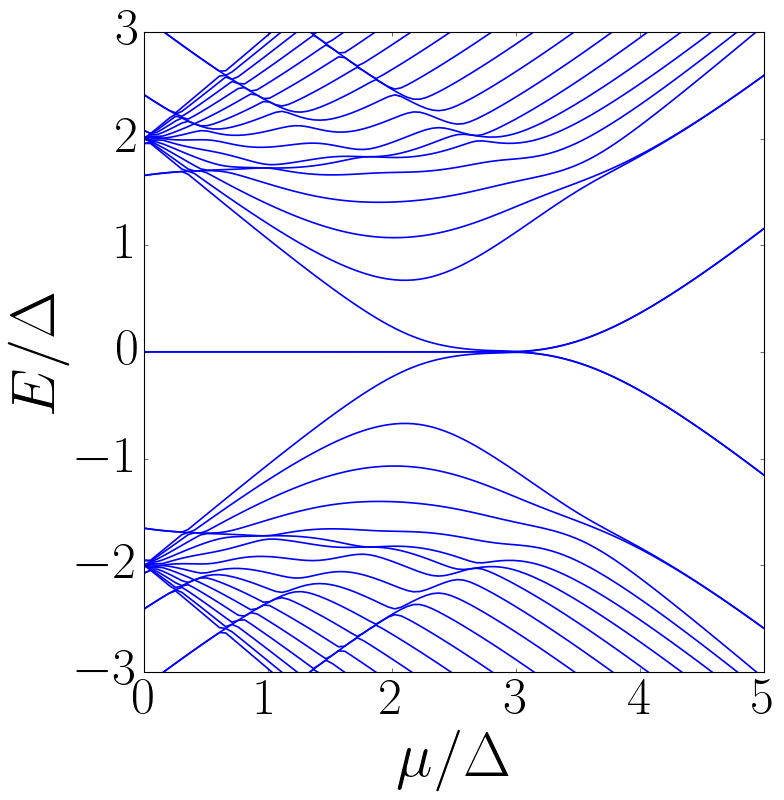}\label{2f}}
	\quad
	\subfigure[]{\includegraphics[width=0.23\textwidth]{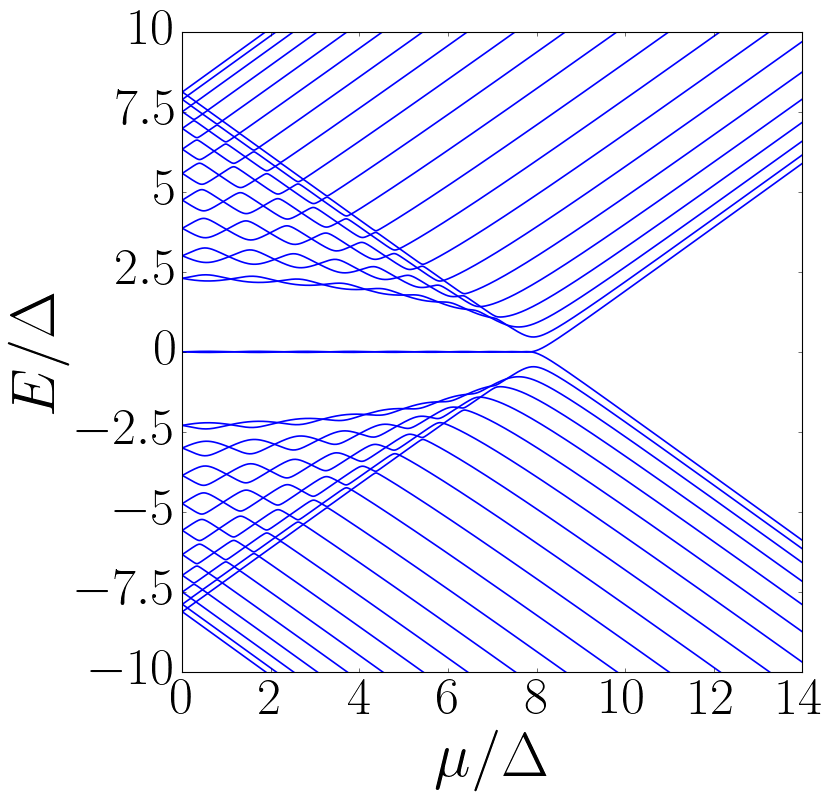}\label{2g}}
	\quad
	\subfigure[]{\includegraphics[width=0.23\textwidth]{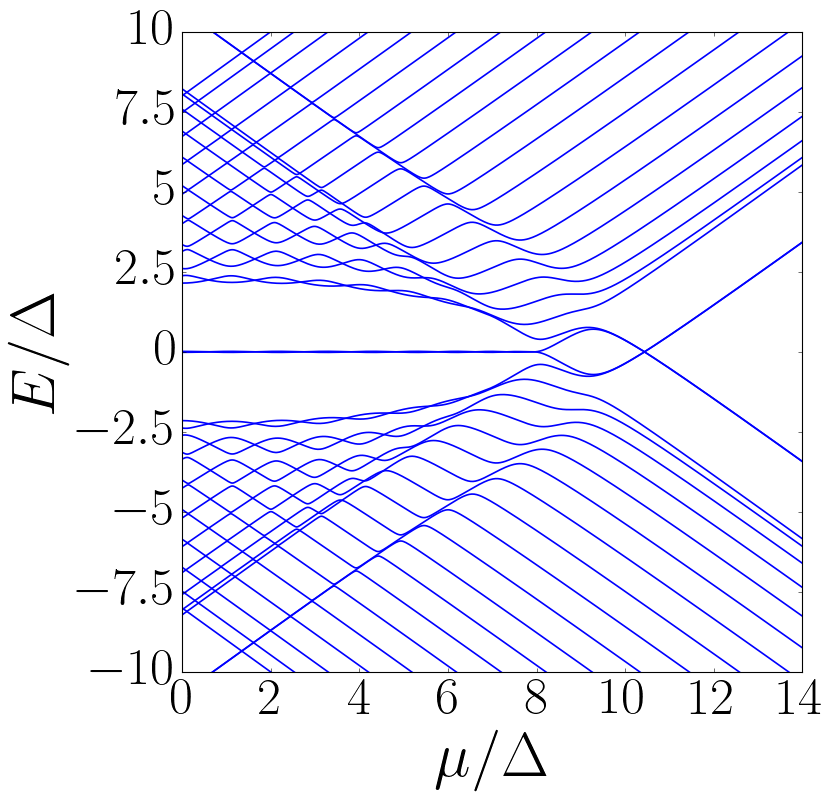}\label{2h}}
	\quad

	\caption{Eigenspectrum as a function of $\mu/\Delta$ for pristine and disordered chains with different system parameters. (a) Eigenspectrum for a pristine setup with N=100 and $t=\Delta$. (b) Eigenspectrum for a disordered setup with N=100 and $t=\Delta$. (c) Eigenspectrum for a pristine setup with N=100 and $t=4.1\Delta$. (d) Eigenspectrum for a disordered setup with N=100 and $t=4.1\Delta$. (e) Eigenspectrum for a pristine setup with N=21 and $t=\Delta$. (f) Eigenspectrum for a disordered setup with N=21 and $t=\Delta$. (g) Eigenspectrum for a pristine setup with N=21 and $t=4.1\Delta$. (h) Eigenspectrum for a disordered setup with N=21 and $t=4.1\Delta$ For the pristine chains, zero energy modes are observed in the topological regime, i.e, $\mu/\Delta < 2t/\Delta$. However, for the disordered chain, zero energy modes are observed both in the topological regime as well as the trivial regime. }
	\label{fig:2}
\end{figure*}
 \indent Non-local correlations related to topological order and the resulting bulk-boundary correspondence can also be suitably quantified using the degree of entanglement \cite{PhysRevB.73.245115,HOLZHEY1994443,PhysRevLett.90.227902,PhysRevLett.109.267203}, quantified by the von Neumann entropy \cite{PhysRevLett.122.210402}. The idea of using this construct is that entanglement in a higher dimensional Hilbert space is reflected via the mixedness of the density matrix in a reduced dimensional Hilbert space. By dividing the composite system into two subsystems, A and B, each of which having similar internal degrees of freedom, we calculate the mixedness of either subsystem using the von Neumann entropy using the partially traced density matrix : $S_{ent}=-\operatorname{Tr}\left(\rho_{A/B}\ln\rho_{A/B} \right)$. As pointed out in \cite{PhysRevB.73.245115}, away from the phase transition point, the entanglement entropy is directly related to the Berry phase of the ground state. Systems with a non-trivial Berry phase of $\pi \times$ (odd integer), host topologically protected boundary modes and have a non-trivial value of entanglement entropy $> \ln2$ per boundary mode. However, for systems with a trivial Berry phase, the entanglement entropy vanishes. At the point of phase transition, the value of entanglement entropy is dictated by the correlation length, which leads to a spike in the magnitude of the entanglement entropy \cite{hegde2021exploring}, making it an ideal metric for gauging the onset of a topological phase transition.\\
\indent  We now present the numerical results performed on a three-terminal N-TS-N system, as shown in Fig. \ref{1a}. The three terminal setup \cite{puglia,Puglia_Cond_Matrix} consists of a grounded superconductor \cite{Hassler} connected to the chain, and two contacts from which independent voltages $(V_L, V_R)$ can be applied. We consider two situations: pristine and disordered, as discussed in previous works \cite{Pan-2019-prb,Pan-2020}. As shown in Fig. \ref{1b}, the pristine setup has a homogenous on-site potential, whereas the disordered setup, as shown in Fig. \ref{1c}, has a local inhomogeneity in the on-site potential at the ends of the TS chain \cite{Pan-2020}. The presence of local inhomogeneity in on-site potential leads to the manifestation of trivial ABSs in the disordered setup. Our calculations reveal that both non-local conductance and the entanglement entropy provide a clear phase transition signature for a long-enough chain. However, the non-local conductance fails to signal a phase transition for smaller TS chain sizes, whereas the entanglement entropy stays robust to chain size.\\
{\it{Energy Spectra:}} We begin by presenting the equilibrium energy spectra for the pristine and disordered topological superconductor setup. We chose the Kitaev chain, a archetypical topological superconductor as our choice for the TS chain. The energy spectrum for the Kitaev chain is well known in literature and is obtained by numerically diagonalizing the tight-binding Hamiltonian:
\begin{equation}
\hat{H}=- \sum_{i=1}^{N}(\mu + V(i) )c_{i}^{\dagger} c_{i}+\sum_{i=1}^{N-1}\left(\Delta c_{i}^{\dagger} c_{i+1}^{\dagger}-t c_{i+1}^{\dagger} c_{i}+\text {h.c.}\right),
\end{equation}
where, $c_{i}^{(\dagger)}$ represents the annihilation (creation) operator on a site i, $\mu$ is the electrochemical potential for the setup which remains constant throughout the chain, $V(i)$ is the additional electrostatic potential at the $i$th site, $t$ is the nearest neighbor inter-site hopping term, h.c., stands for the hermitian conjugate, and $\Delta$ is the superconducting pairing term coupling two neighboring sites as dictated by the Kitaev chain model. \\
\indent In the topological phase space, the region $\mu/\Delta < 2t/\Delta$ represents the topological phase and the region $\mu/\Delta > 2t/\Delta$ represents the trivial phase. Figure \ref{fig:2} shows the eigenspectrum for different setup configurations as a function of $\mu$ with constant $t$ and $\Delta$. Figure \ref{2a} shows the eigenspectrum for a pristine chain of length $N=100$ with $t=\Delta$, which corresponds to the Kitaev point \cite{Leumer_2020,leumer2020linear}. As expected, the topological phase ($\mu/\Delta < 2 $) hosts two MZMs and in the trivial phase ($\mu/\Delta > 2$) the MZMs vanish. However, for a disordered chain of length $N=100$ with $t=\Delta$, as shown in Fig. \ref{2b}, zero-energy modes are observed in the trivial regime as well. The figure clearly shows two zero-energy modes for $\mu/t > 2$, which is within the trivial phase. These zero-energy modes are disorder-induced ABSs \cite{Sci_Post_Wimmer}. \\
\indent Figure \ref{2c} depicts the eigenspectrum for a pristine chain of length $N=100$ with $t=4.1\Delta$, where the system hosts two MZMs in the topological regime ($\mu/\Delta < 8.2$). For a disordered setup, as shown in Fig. \ref{2d}, the system hosts near-zero energy modes in the trivial regime as well. We observe similar features for the systems with smaller chain size of $N=21$ as shown in Figs. \ref{2e}-\ref{2h}. The disordered setups host near-zero energy ABSs in the trivial regime and true MZMs in the topological regime. From these results, we can conclude that the presence of disorder induces zero energy modes in the trivial regime. \\
{\it{Conductance Matrix :}} We now present the relation for the conductance matrix for the setup. As shown in Fig. \ref{1a}, we apply separate voltages $V_{L(R)}$ and measure terminal currents $I_{L(R)}$ at the left and the right contacts, respectively.  Before defining the conductance matrix ($[G]$), we need the relation for the terminal currents $I_{L}$ and $I_{R}$.\\
\indent We use the Keldysh non-equilibrium Green's function formalism to evaluate the terminal currents \cite{leumer2020linear,Duse_2021}. The Keldysh Green's functions are defined over the Keldysh contour, which typically involves the retarded (advanced) Green's function $G^{r(a)}$ and the lesser (greater) Green's function $G^{<(>)}$. The Fourier transform of the Green's function into the energy domain gives the standard steady-state prescription for the transport calculations \cite{leumer2020linear,Duse_2021} involving the density of states, the correlators, and the currents across the setup. Details of this calculation are presented in the supplementary material. In the wide band approximation limit, $G^{<}$ is calculated using the contact broadening matrices and the contact Fermi energies. Once we have the Green's functions, we can derive the terminal currents using the Keldysh current operator \cite{Datta}. For instance, the terminal electron current at the left contact can be derived (see supplementary material) in the Landauer B\"{u}ttiker form as:
\begin{equation} \label{2}
\begin{aligned}
I_{L}^{(e)}=&-\frac{e}{h} \left\{\int d E T_{A}^{(e)}(E)\left[f\left(E-e V_{L}\right)-f\left(E+e V_{L}\right)\right]\right.\\
&+\int d E T_{C A R}^{(e)}(E)\left[f\left(E-e V_{L}\right)-f\left(E+e V_{R}\right)\right] \\
&\left.+\int d E T_{D}^{(e)}(E)\left[f\left(E-e V_{L}\right)-f\left(E-e V_{R}\right)\right]\right\},
\end{aligned}
\end{equation}
where, $T^{(e)}_{D}(E)$, $T^{(e)}_{A}(E)$, and $T^{(e)}_{CAR}(E)$ represent the energy resolved transmission probabilities for the direct, Andreev and crossed-Andreev processes involving the left and right contacts for the electronic sector of the Nambu space.\\

\begin{figure}[!tbp]
	\centering
	\subfigure[]{\includegraphics[width=0.45\textwidth]{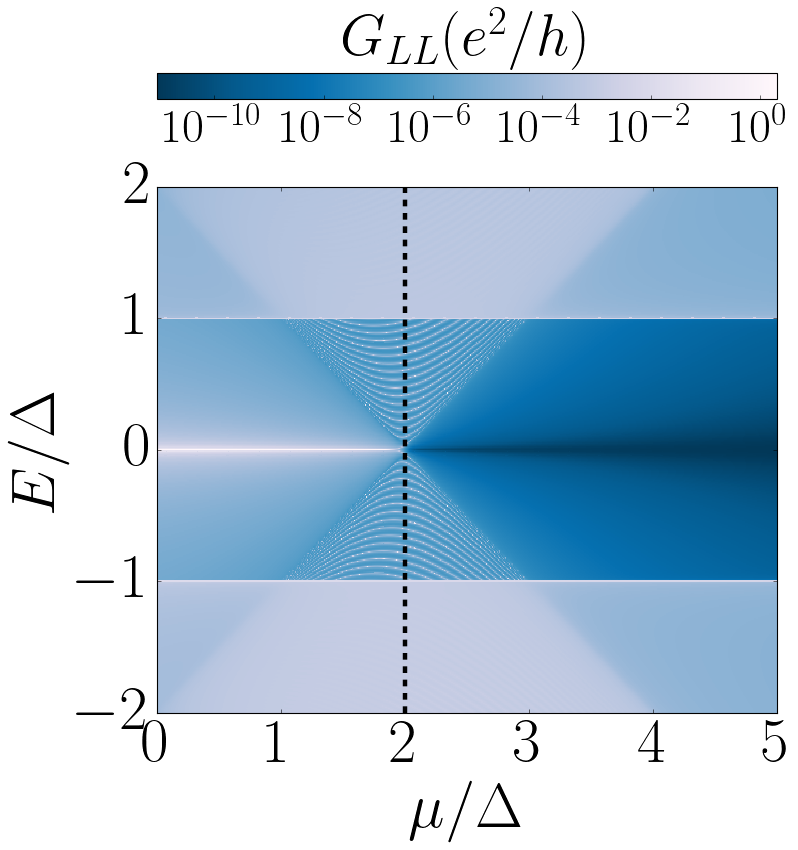}\label{4a}}
	\quad
	\subfigure[]{\includegraphics[width=0.45\textwidth]{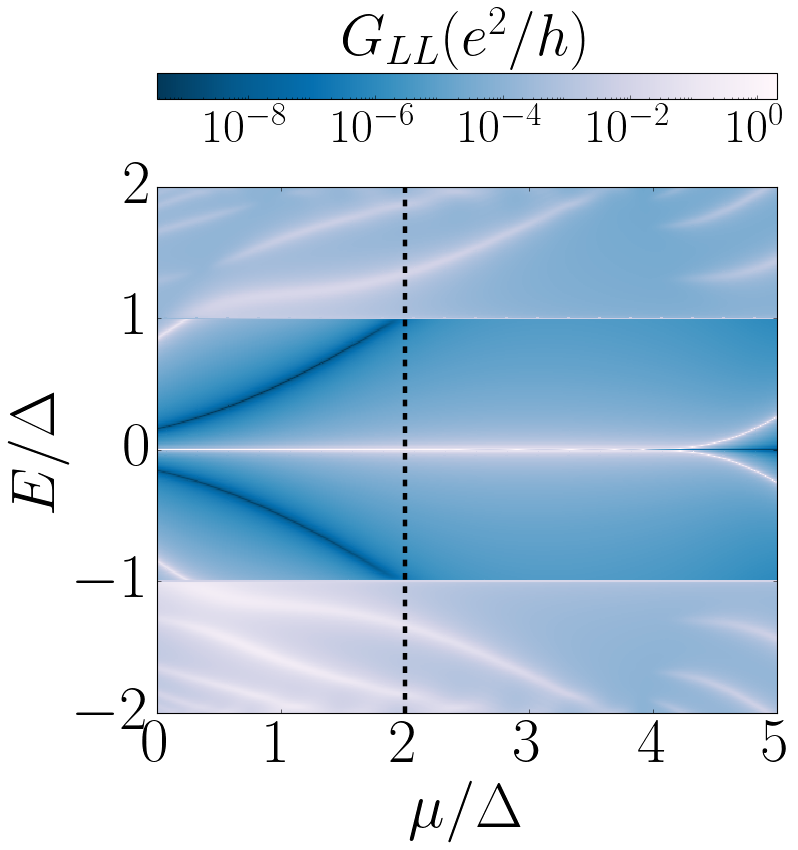}\label{4b}}
	\quad
	\subfigure[]{\includegraphics[width=0.45\textwidth]{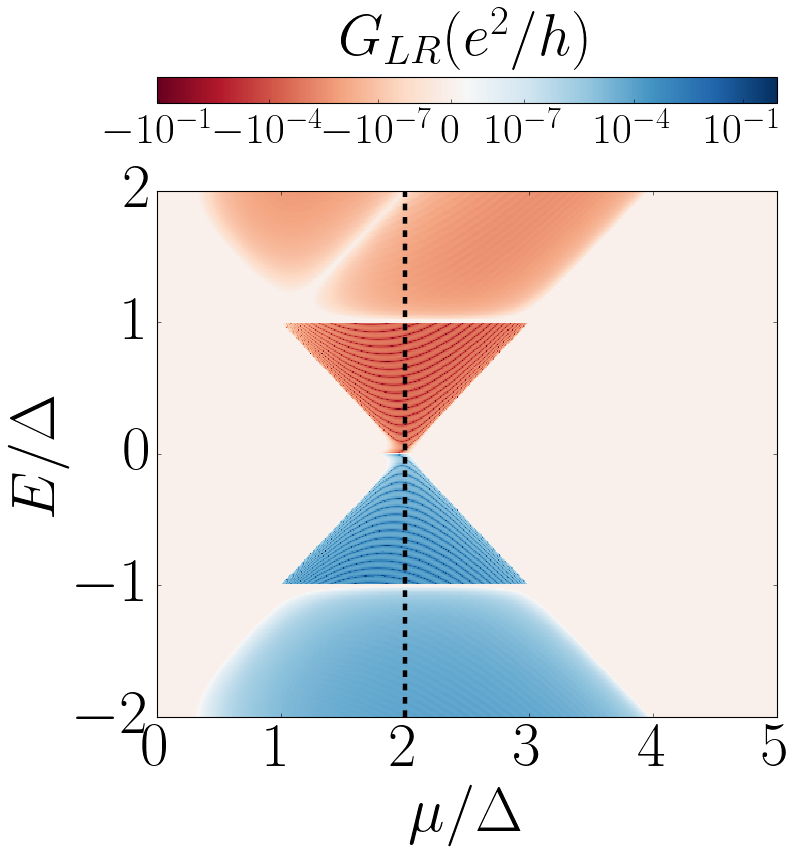}\label{4c}}
	\quad
	\subfigure[]{\includegraphics[width=0.45\textwidth]{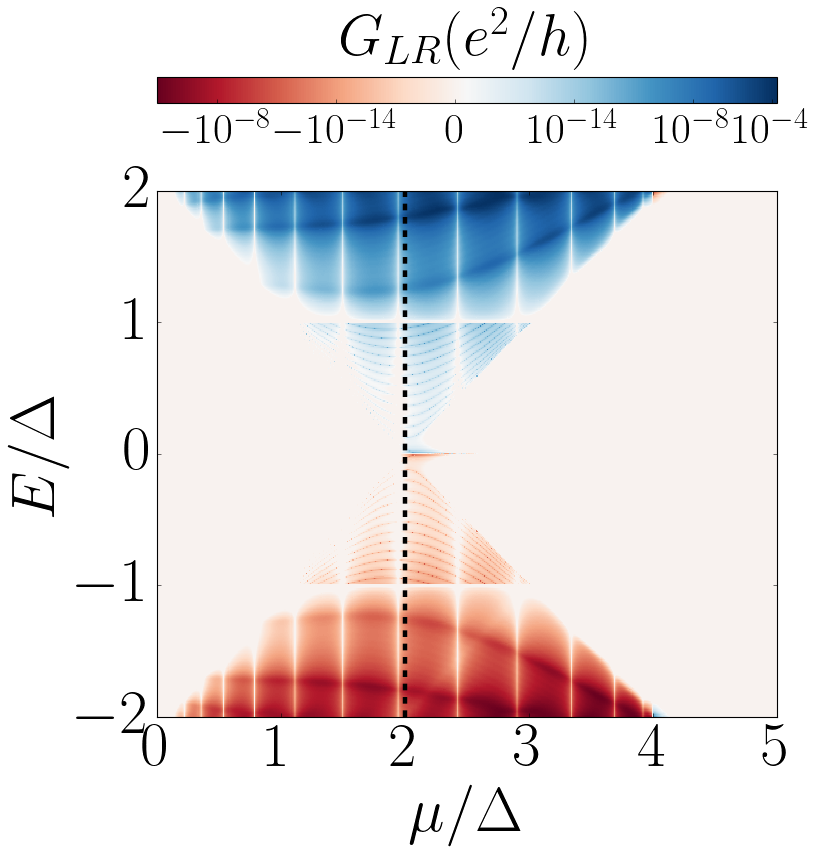}\label{4d}}
	\quad
	\subfigure[]{\includegraphics[width=0.45\textwidth]{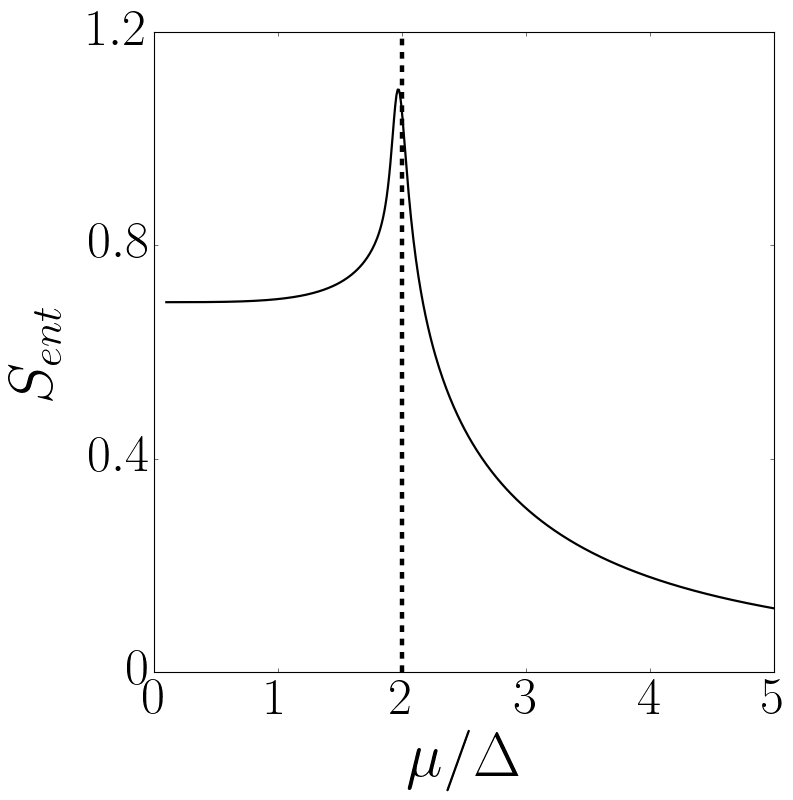}\label{4e}}
	\quad
	\subfigure[]{\includegraphics[width=0.45\textwidth]{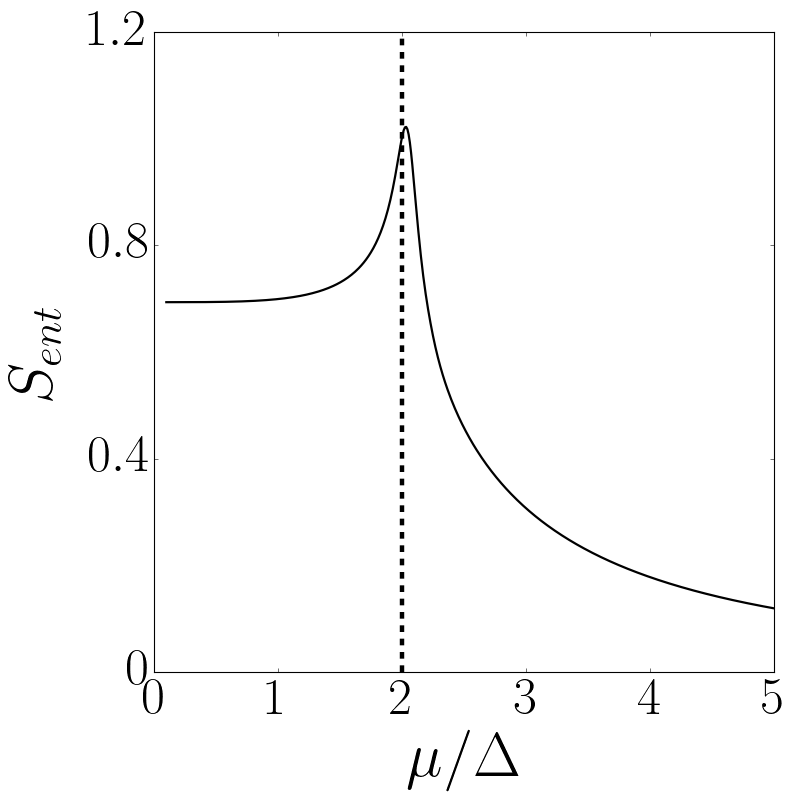}\label{4f}}
	\quad	
	\caption{Results for the setup with N=100 and $t=\Delta$. The black vertical lines represent the topological phase boundary. Local conductance signature for (a) the pristine setup, and (b) for the disordered setup. The ZBCP is observed in the topological regime in (a) and is observed for both the trivial and the topological regimes in (b). Non-local conductance signature for (c) the pristine setup, and (d) the disordered setup. The gap closing is observed at the topological phase boundary in both (c) and (d). Entanglement entropy signatures for (e) the pristine setup, and (f) the disordered setup. Observed features are similar in both setups. The non local conductance shows a gap closing signature at topological phase transition point for both configurations. Thus, both entanglement entropy and non-local conductance can faithfully detect a topological phase transition for the pristine and disordered setups.}
	\label{fig:4}
\end{figure}

\indent Using the terminal currents, the conductance matrix $[G]$ can be defined as:
\begin{equation} \label{eq3}
\mathrm{[G]}=\left(\begin{array}{cc}
G_{L L} & G_{L R} \\
G_{R L} & G_{R R}
\end{array}\right)=\left(\begin{array}{ll}
\left.\frac{\partial I_{L}}{\partial V_{L}}\right|_{V_{R}=0} & \left.\frac{\partial I_{L}}{\partial V_{R}}\right|_{V_{L}=0} \\
\left.\frac{\partial I_{R}}{\partial V_{L}}\right|_{V_{R}=0} & \left.\frac{\partial I_{R}}{\partial V_{R}}\right|_{V_{L}=0}
\end{array}\right),
\end{equation}
where the diagonal matrix elements represent the local conductance at the left and right contacts, whereas the off-diagonal components represent the non-local conductance. \\

\begin{figure}[!htbp]
	\centering
	\subfigure[]{\includegraphics[width=0.45\textwidth]{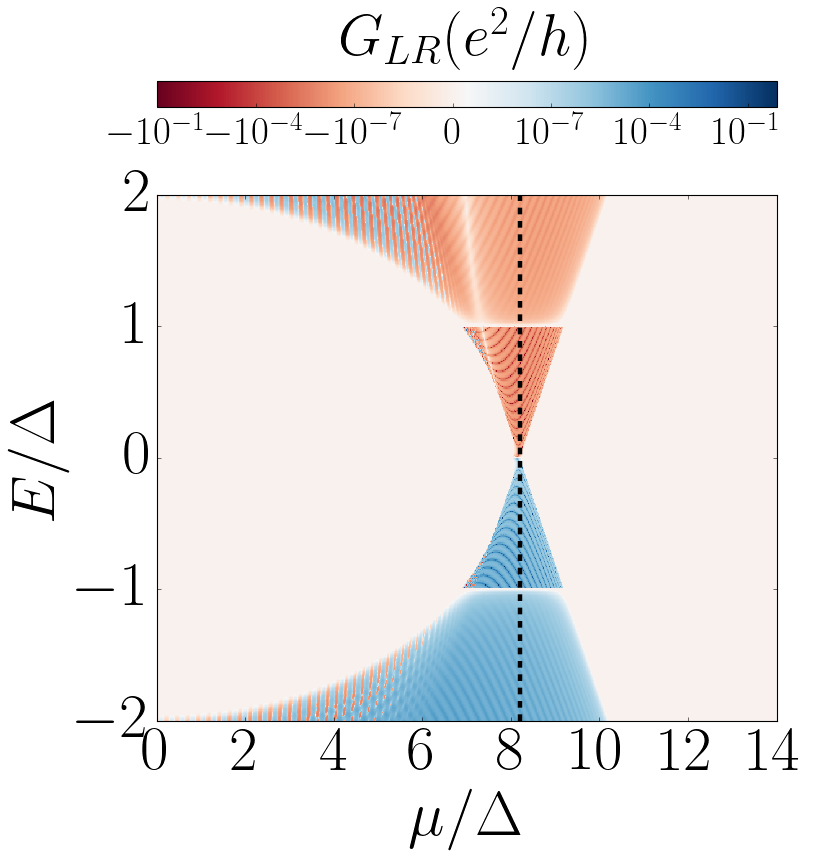}\label{5a}}
	\quad
	\subfigure[]{\includegraphics[width=0.45\textwidth]{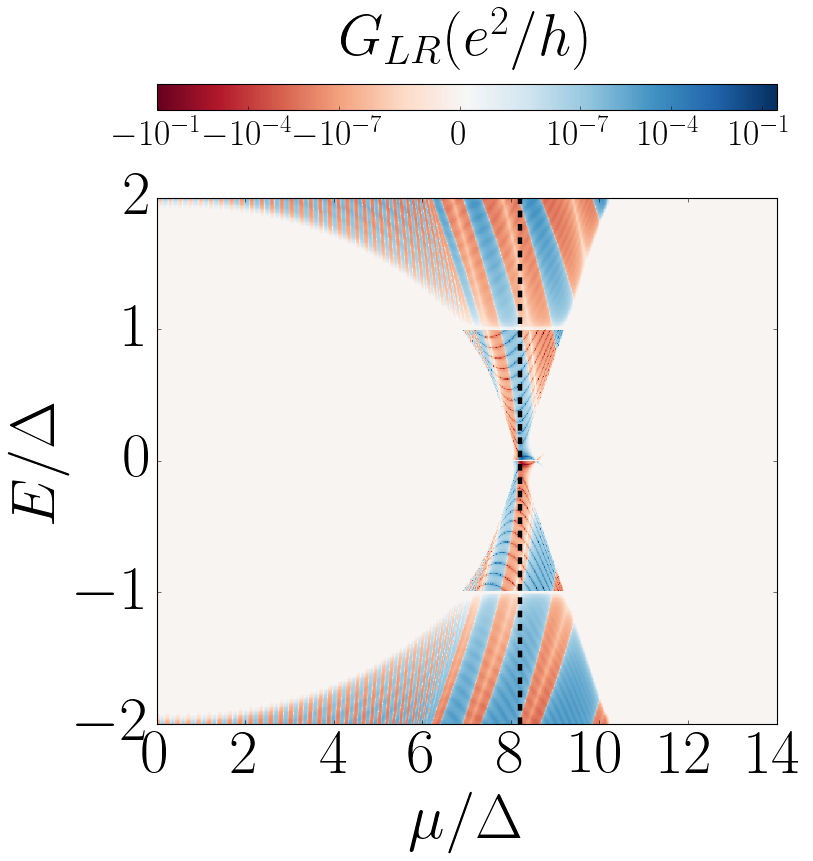}\label{5b}}
	\quad
	\subfigure[]{\includegraphics[width=0.45\textwidth]{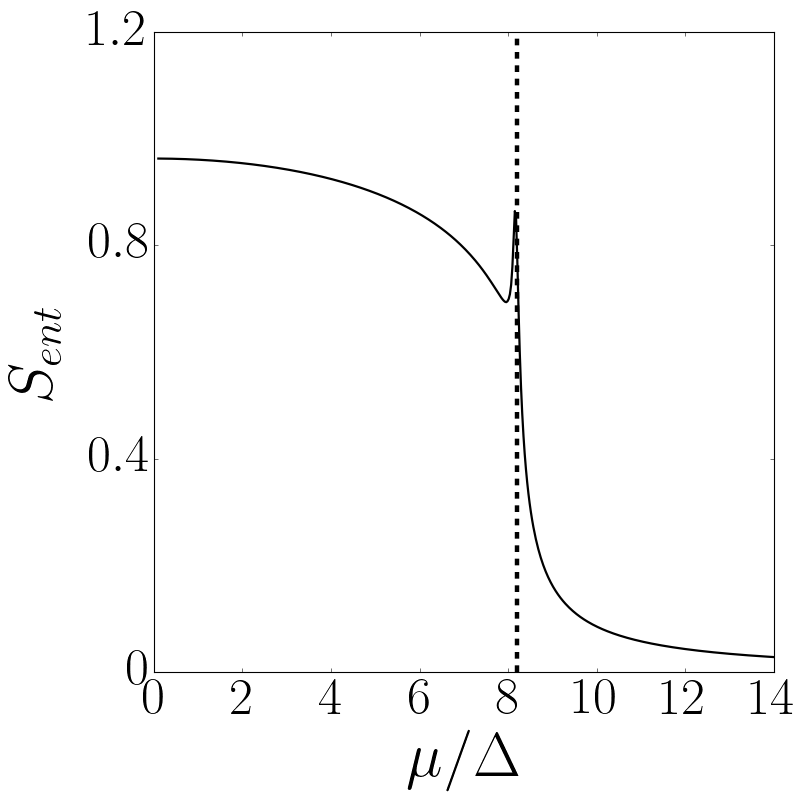}\label{5c}}
	\quad
	\subfigure[]{\includegraphics[width=0.45\textwidth]{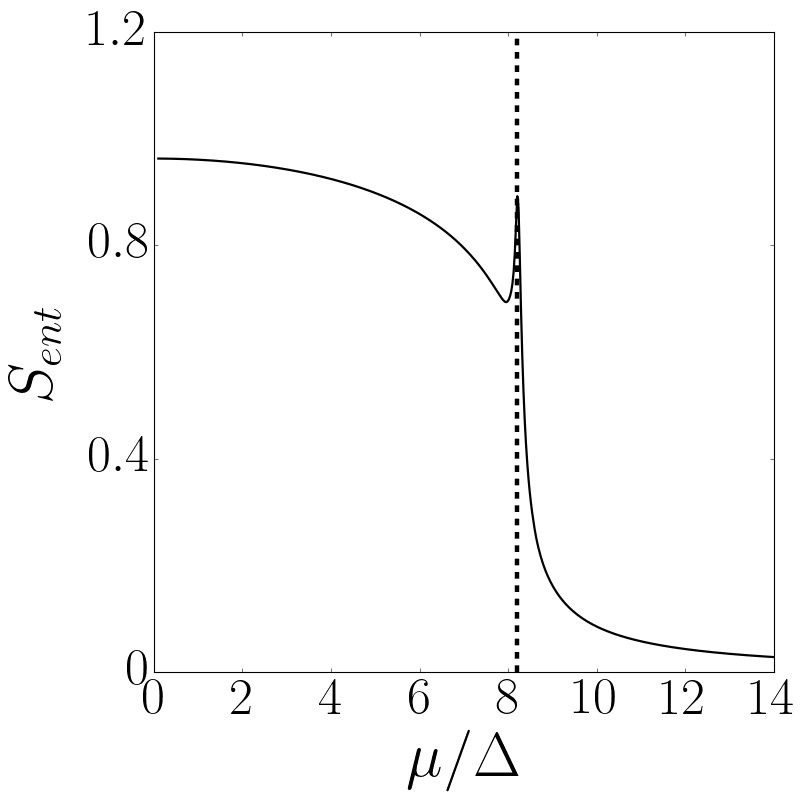}\label{5d}}
	\quad	
	\caption{Results for the setup with N=100 and $t=4.1\Delta$. The black vertical lines represent the topological phase boundary. Non-local conductance signature for (a) the pristine setup, and (b) the disordered setup. The gap closing is observed at the topological phase boundary in both (a) and (b) . Entanglement entropy signatures for (c) the pristine setup, and (d) the disordered setup. The non local conductance shows a gap closing signature at topological phase transition point for both configurations. Observed features are similar in both setups. Thus, both entanglement entropy and non-local conductance can faithfully detect a topological phase transition for the pristine and disordered setups.}
	\label{fig:5}
\end{figure}
\indent The local conductance at the left contact can be derived by taking a partial derivative of the left terminal current ($I_L$), as given in \eqref{2}, with the left contact voltage ($V_L$), and the right contact voltage ($V_R$) set to zero, and is given by : $G_{LL}=\left.\frac{\partial I_{L}}{\partial V_{L}}\right|_{V_{R}=0}$. This translates to the following, in the Landauer-B\"uttiker form:
\begin{equation} \label{4}
\begin{aligned}
\left.G_{LL}(V)\right|_{T \rightarrow 0} \equiv \frac{e^2}{h}\left[T_{A}(E=e V)+T_{A}(E=-e V)\right.+\\
\left.T_{C A R}(E=eV)+T_{D}(E=e V)+G_{LL}^{Gnd}(V)\right],
\end{aligned}
\end{equation}
where the last term $G_{LL}^{Gnd}(V)$ comes as an additional factor due to currents flowing into the grounded terminal (see supplementary material).\\

\begin{figure}[!htbp]
	\centering
	\subfigure[]{\includegraphics[width=0.45\textwidth]{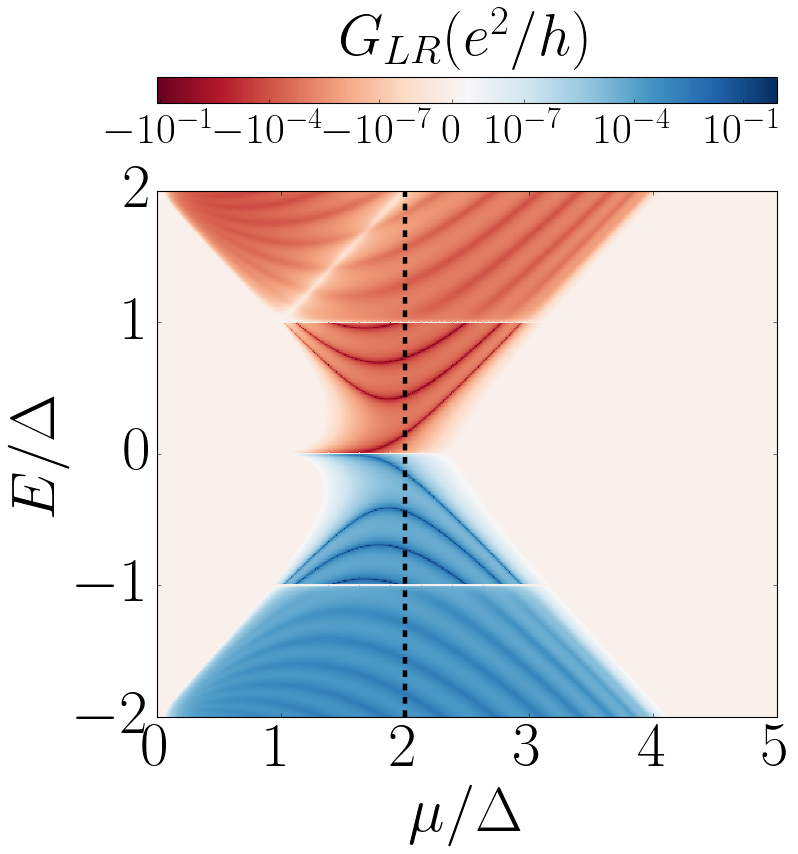}\label{6a}}
	\quad
	\subfigure[]{\includegraphics[width=0.45\textwidth]{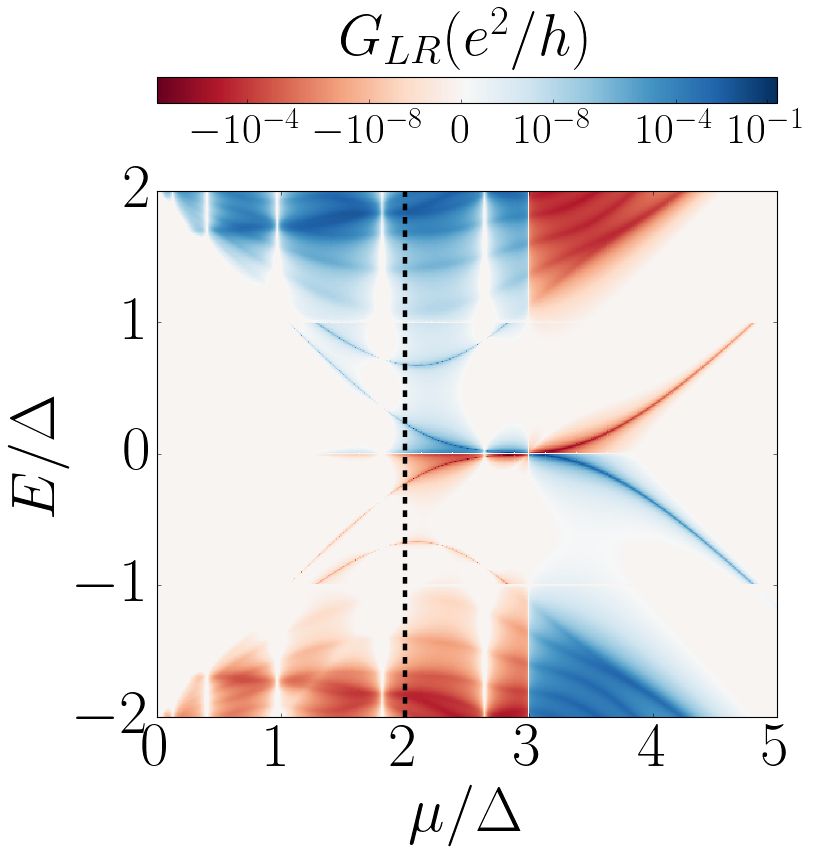}\label{6b}}
	\quad
	\subfigure[]{\includegraphics[width=0.45\textwidth]{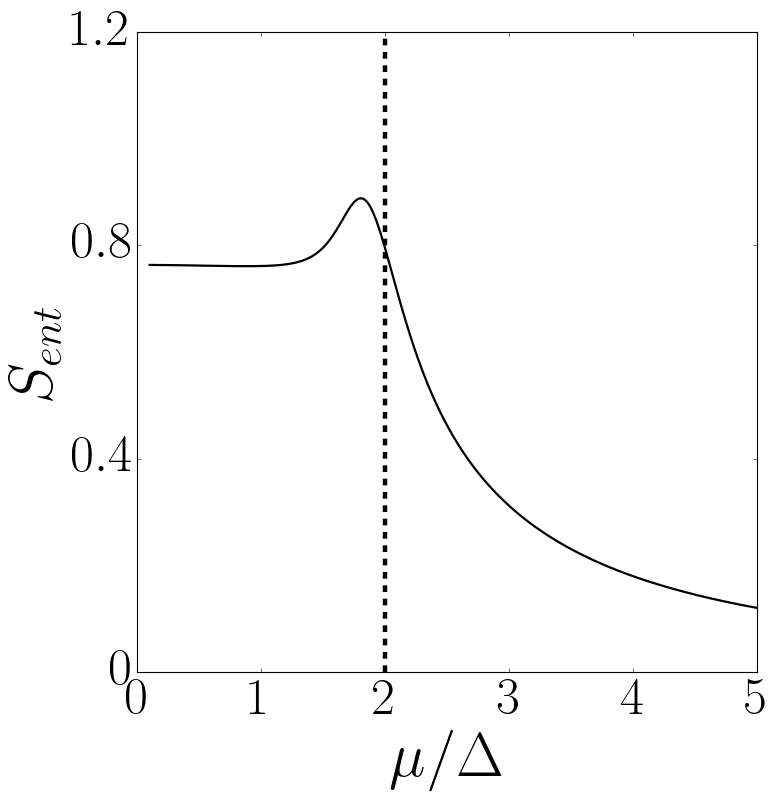}\label{6c}}
	\quad
	\subfigure[]{\includegraphics[width=0.45\textwidth]{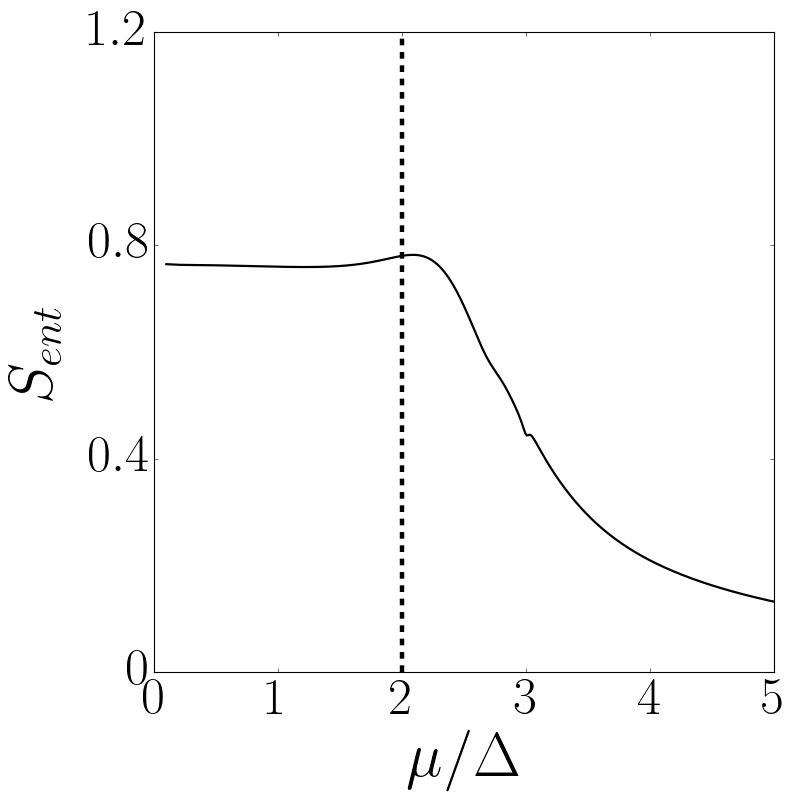}\label{6d}}
	\quad	
	\caption{Conductance matrix and entanglement entropy for the setup with N=21 and $t=\Delta$. The black vertical lines represent the topological phase boundary. Non-local conductance signatures for (a) the pristine setup, and (b) for the disordered setup. The gap closing is observed at the topological phase boundary in (a) and is observed before the topological phase transition in (b). Entanglement entropy signatures for (c) the pristine setup, and (d) for the disordered setup. The observed entanglement entropy features are similar in both setups. The non-local conductance gap closing occurs before the topological phase transition for the disordered setup. Thus, entanglement entropy can faithfully detect a topological phase transition for both pristine and disordered setups whereas non-local conductance fails to do so. }
	\label{fig:6}
\end{figure}

\indent The non-local conductance formula can be derived by taking a partial derivative of the left terminal current ($I_L$), as given in \eqref{2}, over the right terminal voltage ($V_R$), with the left terminal voltage ($V_L$) set to zero, such that, $G_{LR}=\left.\frac{\partial I_{L}}{\partial V_{R}}\right|_{V_{L}=0}$. This gives us the non-local conductance formula in terms of the transmission probabilities as:
\begin{equation} \label{eq5}
\begin{aligned} 
\left.G_{LR}(V)\right|_{T \rightarrow 0} \equiv \frac{e^2}{h}\left[T_{D}(E=e V)-T_{CAR}(E=-e V)\right],
\end{aligned}
\end{equation}
from which, we note that the non-local conductance can show negative values \cite{Akhmerov} that can be used to signal a topological phase transition. \\
{\it{Entanglement Entropy :}} We finally define the entanglement entropy metric for the TS chain. It is important to note that the entanglement entropy is calculated for the isolated TS chain. To calculate the entanglement entropy, we first evaluate the correlation matrix for the chain. The correlation matrix ($C$) is defined as:
\begin{equation} \label{6}
C_{n m}=\left(\begin{array}{cc}
\left\langle c^{\dagger} c\right\rangle & \left\langle c^{\dagger} c^{\dagger}\right\rangle \\
\langle c c\rangle & \left\langle c c^{\dagger}\right\rangle
\end{array}\right)_{n m},
\end{equation}
where the matrix elements of the correlation matrix can be calculated by evaluating the expectation values of the operators quadratic in $c$, $c^{\dagger}$, in the BCS ground state. Once we evaluate the correlation matrix, we partition the chain into two subsystems, A and B, and assess the truncated correlation matrix for either of the subsystems. The matrix elements for the truncated correlation matrix ($\tilde C$) for subsystem $A$ can be defined as follows:
\begin{equation}
\begin{aligned}
\tilde{C}_{n m} &= C_{(n,m) \in A}.
\end{aligned}
\end{equation}
Following the calculation detailed in the supplementary material \cite{PhysRevB.64.064412}, we can prove that the entanglement entropy and the eigenvalues of the truncated correlation matrix are related through the formula:
\begin{equation} \label{8}
\begin{aligned}
S_{ent} =-\sum_{k=1}^{M}\left[\xi_{k} \log \xi_{k}+\left(1-\xi_{k}\right) \log \left(1-\xi_{k}\right)\right],
\end{aligned}
\end{equation}
where $\xi_{k}$ denote the eigenvalues of the truncated correlation matrix.

\begin{figure}[!tbp]
	\centering
	\subfigure[]{\includegraphics[width=0.45\textwidth]{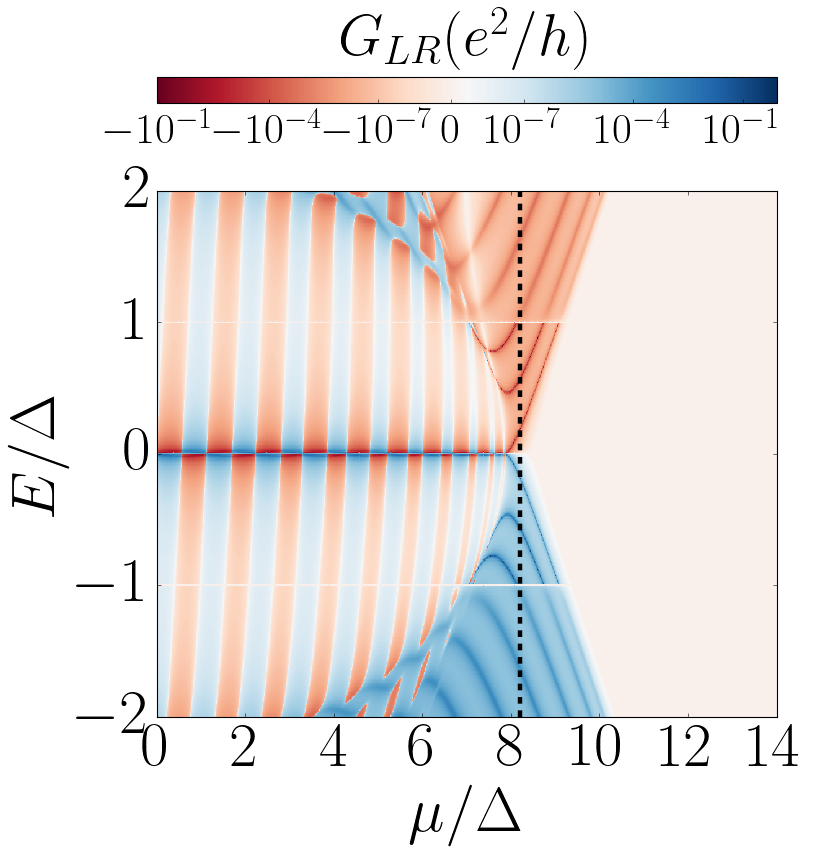}\label{7a}}
	\quad
	\subfigure[]{\includegraphics[width=0.45\textwidth]{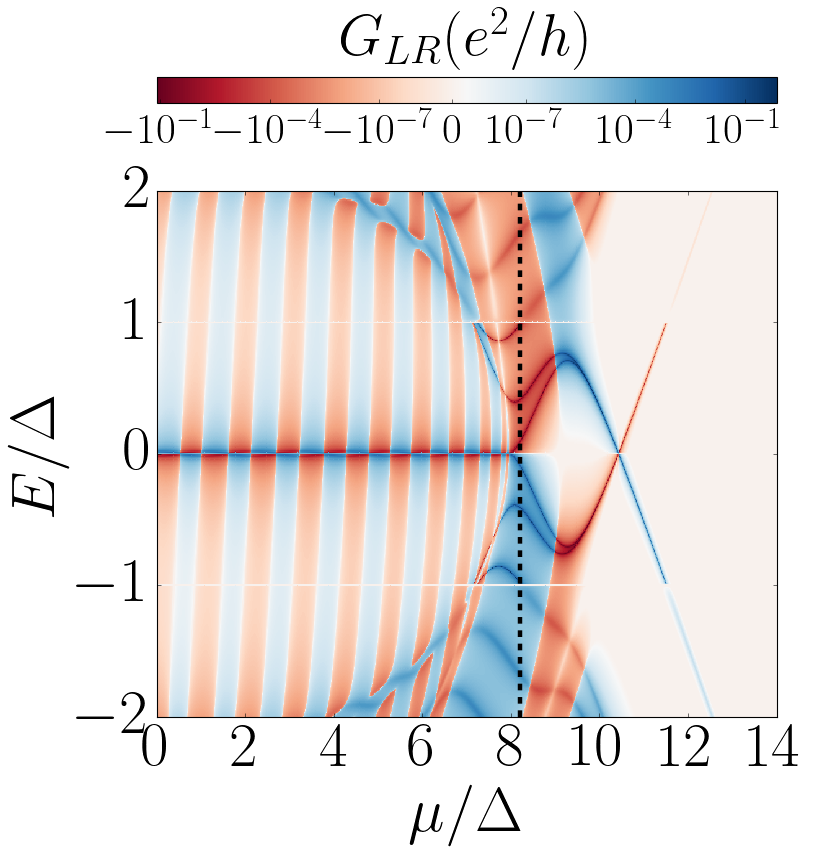}\label{7b}}
	\quad
	\subfigure[]{\includegraphics[width=0.45\textwidth]{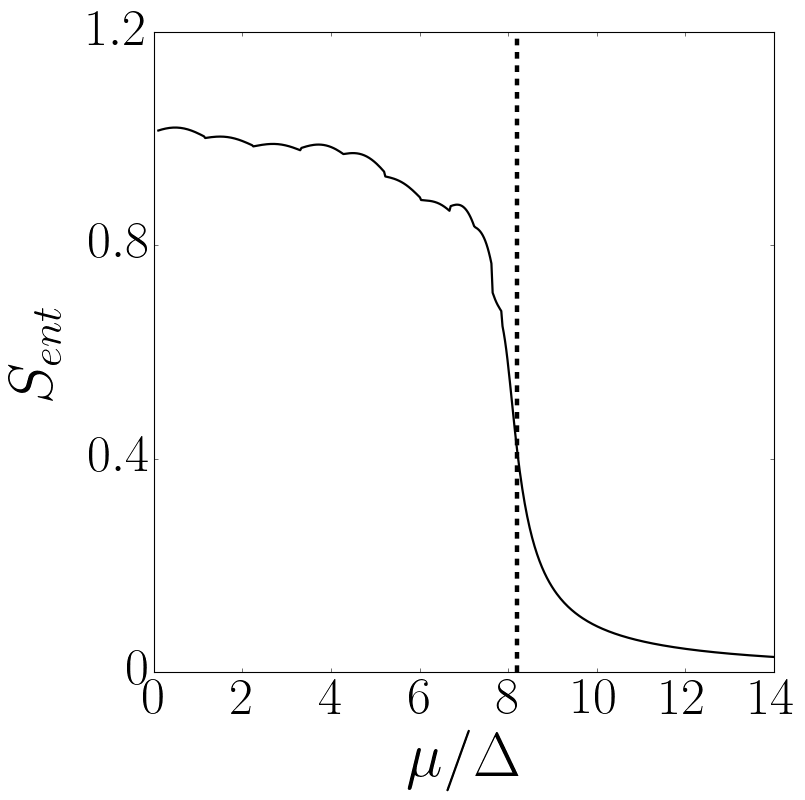}\label{7c}}
	\quad
	\subfigure[]{\includegraphics[width=0.45\textwidth]{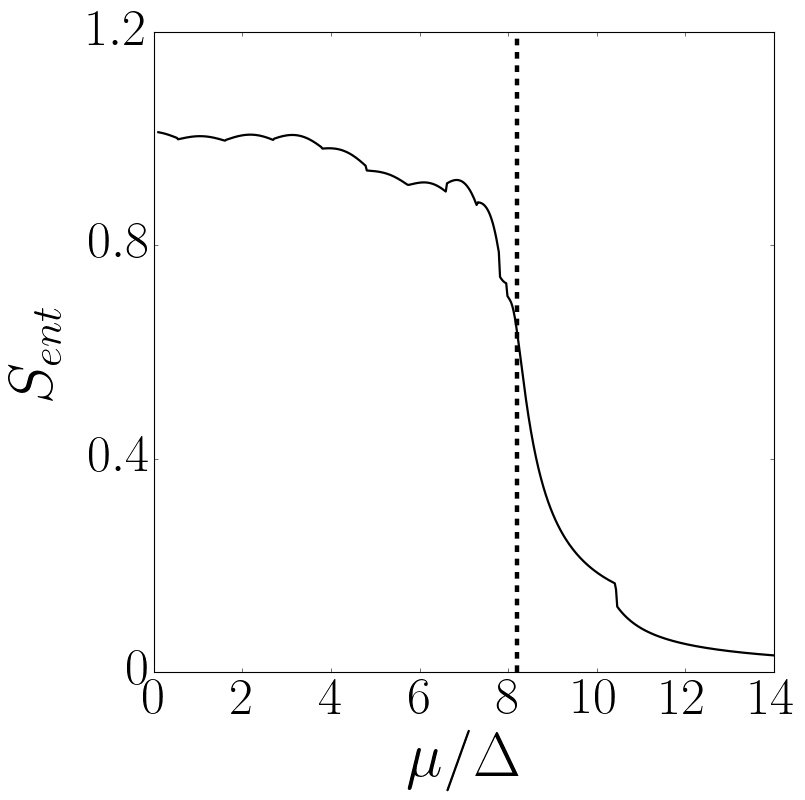}\label{7d}}
	\quad	
	\caption{Conductance matrix and entanglement entropy for the setup with N=21 and $t=4.1\Delta$. The black vertical lines represent the topological phase boundary. Non-local conductance signature for (a) the pristine setup, and (b) the disordered setup. The gap closing is observed at the topological phase boundary in (a)  and is observed before the topological phase transition in (b) . Entanglement entropy signatures for (c) the pristine setup, and (d) the disordered setup. The observed entanglement entropy features are similar in both setups. The non-local conductance gap closing occurs before the topological phase transition for the disordered setup. Thus, entanglement entropy can faithfully detect a topological phase transition for both pristine and disordered setups whereas non-local conductance fails to do so.}
	\label{fig:7}
\end{figure}
 Now that we have the conductance matrix and entanglement entropy equations, we illustrate the numerical results for our setup for different system parameters. Figure \ref{fig:4} depicts the numerical simulations performed for the Kitaev chain with system size $N=100$ and $t/\Delta=1$ . Using the local conductance formula \eqref{4}, we present the local conductance signatures for the pristine and disordered setups in Figs. \ref{4a} and \ref{4b}. We first re-emphasize the well-known fact that true MBSs show up as ZBCPs with a quantized conductance of $2e^2/h$ in the local conductance spectrum in the tunneling regime. This phenomenon can be clearly observed in Fig. \ref{4a}, where the topological regime ($\mu/\Delta < 2$) exhibits a sharp ZBCP. However, the ZBCP signature is also observed in the trivial regime($\mu/\Delta > 2$) for a disordered setup, as shown in Fig. \ref{4b}. This indicates that disorder-induced ABSs can also give rise to ZBCPs in the local conductance spectrum. \\
\indent Next, we present the non-local conductance signatures for the pristine and disordered setups using \eqref{eq5}. As pointed out in \cite{Akhmerov}, the bulk gap in the non-local conductance spectrum closes at the topological phase transition point irrespective of the setup configuration. For the above choice of system parameters, we clearly observe the gap closing at the topological phase transition point for pristine and disordered setups, as shown in Figs. \ref{4c} and \ref{4d}, and thus we can say that the non-local conductance can faithfully detect a topological phase transition. Figures \ref{4e} and \ref{4f} depict the entanglement entropy spectrum for the pristine and disordered setups, calculated numerically using \eqref{8}. We observe that for both the setup configurations, the entanglement entropy has a non-trivial value in the topological regime and drops to a near-zero value at the topological phase transition point.  The non-trivial value of entanglement entropy is associated with the fact that the topological regime hosts topologically protected MZMs that are non-locally correlated. The correlation length governs the height of the peak at the phase transition point for the setup \cite{PhysRevB.73.245115}. We re-emphasize the fact that the disordered chain hosts near-zero energy quasi-MZMs in the trivial regime. Nevertheless, these quasi-MZMs give a near-zero entanglement entropy value due to their local nature and thus can be clearly distinguished from true MZMs using entanglement entropy. \\
\indent Figure \ref{fig:5} depicts the numerical simulation results for the chain with system size $N=100$ and $t/\Delta=4.1$. An important point to note is that the local conductance signatures give similar features as presented in Figs. \ref{4a} and \ref{4b} irrespective of the system parameters. Thus, for the above system parameters, we directly start with illustrating the non-local conductance signatures. As pointed out in Figs. \ref{5a} and \ref{5b}, we again observe a gap-closing signature at the topological phase transition point irrespective of the presence of disorder. Thus, for the above choice of parameters, the non-local conductance faithfully detects a topological phase transition. Figures \ref{5c} and \ref{5d} present the entanglement entropy signatures for the pristine and disordered configurations. Here, we again observe a non-trivial entanglement entropy value in the topological regime and a near-zero value in the trivial regime for both pristine and disordered setups with the non-trivial value attributed to the non-local correlations for true MZMs.  \\
\indent We now present the calculations performed for a system with a smaller size. In Fig. \ref{fig:6} we illustrate the numerical simulation results for a system with system size $N=21$ and $t/\Delta=1$. Based on the original proposal \cite{Akhmerov}, we again expect the gap in the non-local conductance spectrum to close at the topological phase point. We observe this feature for the pristine chain, as shown in Fig. \ref{6a}. However, contrary to what was shown in \cite{Akhmerov} and what we observed for a larger system size of $N=100$ as depicted in Figs. \ref{fig:4} and \ref{fig:5}, we observe a gap-closing well before the topological phase transition for the disordered setup, as shown in Fig. \ref{6b}. This observation shows that the non-local conductance is prone to the finite size effect and fails to detect a topological phase transition for smaller system sizes. The entanglement entropy spectrum, as shown in Figs. \ref{6c} and \ref{6d}, shows a non-trivial value for the topological regime and a near-zero value for the trivial regime for both pristine and disordered configurations. This observation establishes the robustness of the entanglement entropy metric to system parameters.\\
\indent Finally, Fig. \ref{fig:7} illustrates the numerical simulation results for the chain with system size $N=21$ and $t/\Delta =4.1$. The observed results have similar features as the system with $N=21$ and $t/\Delta=1$, as shown in Fig. \ref{fig:6} . The pristine setup shows a gap-closing signature at the topological phase transition point, as shown in Fig. \ref{7a}. However, for a disordered setup, as shown in Fig. \ref{7b}, we observe the gap closing well before the topological phase transition. This result reiterates that non-local conductance as a metric to detect topological phase transition is prone to the finite-size effect. The entanglement entropy again has a non-trivial value in the topological regime and a near-zero value in the topological regime for both pristine and disordered configurations, as shown in Figs. \ref{7c} and \ref{7d}. This observation further emphasizes the robustness of entanglement entropy as a metric to distinguish between the trivial and topological phases.\\
\indent Experimental non-local conductance measurements for disordered physical systems have signaled a gap-closure signature before the appearance of ZBCPs in the local conductance spectrum \cite{puglia}. This rules out the connection between gap-closure and ABSs. One of the possible reasons for the gap-closure can be related to the characteristic length scales of the disorder-induced subgap states being longer than the typical device lengths. For our device setup with local inhomogeneity, we observe a gap closure in the trivial regime for device lengths smaller than the characteristic localization length of the ABSs. However, the entanglement entropy signature remains robust to the device size and faithfully detects a topological phase transition. \\
\indent The ideas presented here pose the challenge of looking beyond conductance spectra as means to ascertaining the presence of topological MZMs and establishing experimental measurement techniques for the entanglement entropy spectra. One of the possible ways is to figure out the exact ground state density matrix through quantum state tomography. However, quantum state tomography is resource-intensive, with the cost scaling exponentially with the subsystem size. Estimating the statistical correlations in the subsystem through the technique of random measurements provides an alternate route for assessing the entanglement entropy \cite{satzinger2021realizing,PhysRevLett.120.050406,PhysRevLett.108.110503,doi:10.1126/science.aau4963}. The key idea behind this protocol is that the second-order R\'enyi entropies can be quantified using statistical correlations between the outcomes of measurements performed in random bases. Other works have presented the connection between entanglement entropy and the electrical shot noise generated by opening and closing a quantum point contact periodically in time \cite{PhysRevLett.102.100502}. It may also be possible to device setups based on quantum thermodynamics \cite{Sergey_Smirnov_1,Sergey_Smirnov_2} to detect such entropic signatures. \\
{\it{Conclusion:}} To summarize, we theoretically proposed the use of topological entanglement entropy as a metric to distinguish between the topological MZM and the trivial ABS. We clearly demonstrated that under various experimental conditions related to the setup, the non-local conductance measurements could yield false positives and premature gap closures. We showed that while both the entanglement entropy and the non-local conductance exhibit a clear topological phase transition signature for long enough pristine nanowires, non-local conductance fails to signal a topological phase transition for shorter disordered wires. While recent experiments have indeed shown premature gap-closure signatures in the non-local conductance spectra, we believe that the entanglement entropy can indeed signal a genuine transition, regardless of the constituent non-idealities in an experimental situation. Our results thereby point toward furthering the development of experimental techniques beyond conductance measurements to achieve a conclusive detection of Majorana zero modes.  \\
 {\it{Acknowedgements:}} The authors acknowledge Adhip Agarwala for insightful discussions. The research and development work undertaken in the project under the Visvesvaraya Ph.D Scheme of the Ministry of Electronics and Information Technology (MEITY), Government of India, is implemented by Digital India Corporation (formerly Media Lab Asia). This work is also supported by the Science and Engineering Research Board (SERB), Government of India, Grant No. Grant No. STR/2019/000030, the Ministry of Human Resource Development (MHRD), Government of India, Grant No. STARS/APR2019/NS/226/FS under the STARS scheme.
\bibliography{main}

\onecolumngrid
\section*{Supplementary Material}
\subsection{Entanglement Entropy}
\renewcommand{\theequation}{S\arabic{equation}}
\renewcommand{\thefigure}{S\arabic{figure}}
\setcounter{equation}{0}
\setcounter{figure}{0}
\subsubsection{Method 1}
Let's start with considering the system defined by the Hilbert space $\mathcal{H}$. To calculate the entanglement entropy we divide the system into two subsystems $A$ and $B$, such that $\mathcal{H}=\mathcal{H}_A \otimes \mathcal{H}_B$. The entanglement entropy for the entire system is given by the von Neumann entropy of subsystem $A$ or $B$. The von Neumann entropy for a subsystem can be calculated using the reduced density matrix for the subsystem which is evaluated as:
\begin{equation} \label{a1}
    S = -\operatorname{Tr} \left[\rho_A\ln{\rho_A}\right] = -\sum_{k} \alpha_{k}\ln{\alpha_{k}},
\end{equation}
where $\alpha_{k}$ represents the $k^{th}$ eigenvalue of the reduced density matrix $\rho_A$. Now, we need to know what is the reduced density matrix $\rho_A$ for our problem and then we need a way to find its eigenvalues. 

To find $\rho_A$, we first need to find the ground state wavefunction $|\Phi_0\rangle$ for the entire chain. Starting with the general form of a Hamiltonian quadratic in the fermionic operators we can write:
\begin{equation} \label{a2}
\hat{H}=\sum_{i, j}^{N}\left[c_{i}^{\dagger} A_{i j} c_{j}+\frac{1}{2}\left(c_{i}^{\dagger} B_{i j} c_{j}^{\dagger}-c_{i} B_{i j} c_{j}\right)\right].
\end{equation}
For the Kitaev chain setup, the matrix $A$ and $B$ are given by:

\begin{equation}
A =\left[\begin{array}{ccccccc}
-\mu & -t & & & & & \\
-t & -\mu & -t & & & & \\
& -t & -\mu & -t & & & \\
& & \ddots & \ddots & \ddots & & \\
& & & -t & -\mu & -t & \\
& & & & -t & -\mu & -t \\
& & & & & -t & -\mu
\end{array}\right] \mspace{20mu} \quad B =\left[\begin{array}{ccccccc}
0 & \Delta & & & & & \\
-\Delta & 0 & \Delta & & & & \\
& -\Delta & 0 & \Delta & & & \\
& & \ddots & \ddots & \ddots & & \\
& & & -\Delta & 0 & \Delta & \\
& & & & -\Delta & 0 & \Delta \\
& & & & & -\Delta & 0
\end{array}\right].
\end{equation}
Although this Hamiltonian does not conserve particle number, it does conserve parity. We use the Bugoliubov transformation to get the eigenstates of the Hamiltonian. This technique uses matrices $g$ and $h$ such that the Hamiltonian is diagonalized:

\begin{equation} \label{a5}
\begin{aligned}
\eta_{k} =\sum_{j} g_{k j} c_{j}+h_{k j} c_{j}^{\dagger} , \mspace{30mu}
\eta_{k}^{\dagger} =\sum_{j} g_{k j} c_{j}^{\dagger}+h_{k j} c_{j},
\end{aligned}
\end{equation}
such that,
\begin{equation} \label{a6}
\hat{H}=\sum_{k} \Lambda_{k} \eta_{k}^{\dagger} \eta_{k}+\mathrm{constant}.
 \end{equation}
The matrices $g$ and $h$ can be chosen using the condition that $\eta_{k}$ and $\eta_{k}^{\dagger}$ anti-commute. The anti-commutation relation imposes the condition:
 \begin{equation} \label{a7}
\left[\eta_{k}, \hat{H}\right]=\Lambda_{k} \eta_{k}.
\end{equation}
Substituting (\ref{a5}) and (\ref{a6}) to (\ref{a7}) imposes constraints on $g$ and $h$ that can be written as:
\begin{equation} \label{a8}
\begin{array}{l}
\sum_{j}\left(g_{k j} A_{j n}-h_{k j} B_{j n}\right)=\Lambda_{k} g_{k n} \\
\sum_{j}\left(g_{k j} B_{j n}-h_{k j} A_{j n}\right)=\Lambda_{k} h_{k n}.
\end{array}
\end{equation}
To solve for $g$ and $h$, two new matrices are defined,
\begin{equation} \label{a9}
\begin{aligned}
\Phi=g+h, \mspace{50mu}
\Psi=g-h.
\end{aligned}
\end{equation}
This reduces (\ref{a8}) to an eigenvalue problem in $\Phi$ and $\Psi$.
\begin{equation} \label{a10}
\begin{array}{l}
\Phi(A-B)(A+B)=\Lambda^{2} \Phi \\
\Psi(A+B)(A-B)=\Lambda^{2} \Psi.
\end{array}
\end{equation}
The eigenvectors for the above eigenvalue problem give us $\Phi$ and $\Psi$. We then use (\ref{a9}) to get back the matrices $g$ and $h$ and thus we have successfully found a way to diagonalize $\hat{H}$. Since $\hat{H}$ conserves parity, its ground state can either have odd or even number of fermions. Without loss of generality, we can assume the ground state to have even number of fermions. The general form of ground state for a Hamiltonian quadratic in fermionic operators and with definite parity is given by the BCS wavefunction which represents a superposition of either odd or even number of particles depending upon the parity of the system,  
\begin{equation} \label{a11}
\left|\Phi_{0}\right\rangle=C \exp \left(\frac{1}{2} \sum_{i j} c_{i}^{\dagger} G_{i j} c_{j}^{\dagger}\right)|0\rangle,
\end{equation}
where $|0\rangle$ is the vacuum state of the operators $c_j$. Since the ground state $|\Phi_0\rangle$ is the vacuum state of opertors $\eta_{j}$, $\eta_{j}|\Phi_{0}\rangle=0$. This relation can be used to find the matrix $G$ in terms of $g$ and $h$ and some tedious calculation leads us to the relation: $gG = h$.

Now that we have the ground state wavefunction for the system $|\Phi_{0}\rangle$, the density matrix is given by $\rho = |\Phi_0\rangle\langle\Phi_0| $ which when substituted with (\ref{a11}) yields:
\begin{equation} \label{a12}
\rho=|C|^{2} \exp \left(\frac{1}{2} \sum_{i j} c_{i}^{\dagger} G_{i j} c_{j}^{\dagger}\right)|0\rangle\langle 0| \exp \left(-\frac{1}{2} \sum_{i j} c_{i} G_{i j} c_{j}\right).
\end{equation}
The next step is to get the reduced density matrix for the sub-system A, using $\rho_A = \operatorname{Tr}_{B}(\rho)$. The final step is to directly diagonalize $\rho_A$ to get its eigenvalues. These eigenvalues can be subsituted to (\ref{a1}) to get the entanglement entropy. But this method requires calculating all the intermediate matrices defined above like $g$, $h$, $G$ and can be quite tedious.  

\subsubsection{Method 2}

Another way to get the entanglement entropy is to use the correlation matrix for the system. A rigorous calculation using Grassman algebra \cite{PhysRevB.64.064412} simplifies the reduced density matrix derived in (\ref{a12}) to:
\begin{equation} \label{a13}
\rho_A=Kexp(-\mathcal{H}_A),
\end{equation}
where $K=|C|^{2}$ and $\mathcal{H}_A$ is the effective Hamiltonian for the sub-system $A$, which has the same general form as the original Hamiltonian $\hat{H}$, the only difference being the matrices $A$ and $B$. If we could figure out a way to get the eigenvalues of $\mathcal{H}_A$, then we can easily get the eigenvalues of $\rho_A$ and thus the entanglement entropy. 

Let's assume that the subsystem $A$ has $M$ sites and the net system has $N$ sites. The effective Hamiltonian $\mathcal{H}_A$ can be written as:
\begin{equation} \label{a14}
\mathcal{H}_A=\sum_{i,j=1}^{2M} \tilde{\mathcal{C}}_i^{\dagger}H_{ij}\tilde{\mathcal{C}}_j,
\end{equation}
where $\tilde{\mathcal{C^\dagger}}=(c_1^{\dagger},...,c_M^{\dagger},c_1,...,c_M )$. $\mathcal{H}_A$ can now be diagonalized by defining $\Psi^{\dagger}=\tilde{\mathcal{C^\dagger}}S=(a_M,...,a_1,a_1^{\dagger},...,a_M^{\dagger})$ with $S$ such that $S^{\dagger}HS=D=diag(-\varepsilon_M,...,-\varepsilon_1,\varepsilon_1,...,\varepsilon_M)$. This simple substitution yeilds:
\begin{equation} \label{a15}
\mathcal{H}_{A}=\sum_{i j=1}^{2 M} \tilde{\mathcal{C}}_{i}^{\dagger} H_{i j} \tilde{\mathcal{C}}_{j}=\sum_{j=1}^{2 M} \Psi_{j}^{\dagger} D_{j j} \Psi_{j}=\sum_{j=1}^{M} a_{j}^{\dagger} \varepsilon_{j} a_{j}-a_{j} \varepsilon_{j} a_{j}^{\dagger}=2 \sum_{j=1}^{M} \varepsilon_{j} a_{j}^{\dagger} a_{j}-E_{0},
\end{equation}
where $E_0=\sum_{j=1}^{M} \varepsilon_{j}$. Since the matrix $H$ is not known, an alternate method is required to find the eigenvalues $\varepsilon$. This is where the concept of correlation matrix comes in. We'll see later in this section how the eigenvalues of the truncated correlation matrix are related to the eigenvalues of $\mathcal{H}_A$. But first, we start by defining the correlation matrix and the truncated correlation matrix and describe how it's calculated.

The correlation matrix for the entire system with size $N$ is defined as:
\begin{equation} \label{a16}
C_{n m}=\left(\begin{array}{cc}
\left\langle c^{\dagger} c\right\rangle & \left\langle c^{\dagger} c^{\dagger}\right\rangle \\
\langle c c\rangle & \left\langle c c^{\dagger}\right\rangle
\end{array}\right)_{n m}=\left\langle\mathcal{C}_{n}^{\dagger} \mathcal{C}_{m}\right\rangle=\operatorname{Tr}\left[\rho \mathcal{C}_{n}^{\dagger} \mathcal{C}_{m}\right],
\end{equation}
where $\mathcal{C^\dagger}=(c_1^{\dagger},....,c_N^{\dagger},c_1,....,c_N )$. The elements of the correlation matrix have the form $\left\langle c_{n}^{\dagger} c_{m}\right\rangle=\left\langle\Phi_{0}\left|c_{n}^{\dagger} c_{m}\right| \Phi_{0}\right\rangle$. To evaluate these expectation values, we follow a procedure exactly similar to what we did in \eqref{a15}. The Hamiltonian for the net system $\hat{H}$ given by (\ref{a2}) can be written as $\hat{H}=\mathcal{C}^{\dagger}H_{BdG}\mathcal{C}$ where
\begin{equation} \label{a17}
H_{BdG}=\frac{1}{2}\left[\begin{array}{cc}
A & B \\
-B & -A
\end{array}\right].
\end{equation}
Just as (\ref{a14}), we can define $\Psi^{\dagger}=C^{\dagger}S= (a_N,...,a_1,a_1^{\dagger},...,a_N^{\dagger})$ to diagonalize $H_{BdG}$, and we finally get:
\begin{equation} \label{a18}
\begin{aligned}
\left\langle\Phi_{0}\left|c_{n}^{\dagger} c_{m}\right| \Phi_{0}\right\rangle &=\sum_{k l}\left\langle\Phi_{0}\left|\Psi_{k}^{\dagger} S_{k n}^{T} S_{m l} \Psi_{l}\right| \Phi_{0}\right\rangle \\
&=\sum_{k l \leq N}\left\langle\Phi_{0}\left|a_{N+1-k} S_{k n}^{T} S_{m l} a_{N+1-l}^{+}\right| \Phi_{0}\right\rangle \\
&=\sum_{k=1}^{N} S_{k, n}^{T} S_{m, k}.
\end{aligned}
\end{equation}
Similarly we can get the other terms:
\begin{equation} \label{a19}
\begin{aligned}
\left\langle\Phi_{0}\left|c_{n} c_{m}\right| \Phi_{0}\right\rangle =\sum_{k>N, l \leq N} S_{n, k} S_{m, l}\left\langle\Phi_{0}\left|a_{k-N} a_{N+1-l}^{\dagger}\right| \Phi_{0}\right\rangle 
=\sum_{k=1}^{N} S_{2N+1-k, n}^{T} S_{m,k},
\end{aligned}
\end{equation}
and 
\begin{equation} \label{a20}
\begin{aligned}
\left\langle\Phi_{0}\left|c_{n}^{\dagger} c_{m}^{\dagger}\right| \Phi_{0}\right\rangle =
\sum_{k=1}^{N} S_{k, n}^{T} S_{m,2N+1-k} \mspace{50mu}
\left\langle\Phi_{0}\left|c_{n} c_{m}^{\dagger}\right| \Phi_{0}\right\rangle &=
\sum_{k=1}^{N} S_{k+N, n}^{T} S_{m, k+N}.
\end{aligned}
\end{equation}
The above relations give us the entire correlation matrix $C$. The truncated correlation matrix $\tilde{C}$ is a sub-matrix of the net correlation matrix $C$ which is defined as follows:
\begin{equation} \label{a22}
\tilde{C}_{n m}=\left(\begin{array}{cc}
\left\langle c^{\dagger} c\right\rangle & \left\langle c^{\dagger} c^{\dagger}\right\rangle \\
\langle c c\rangle & \left\langle c c^{\dagger}\right\rangle
\end{array}\right)_{n, m \in A}=\operatorname{Tr}\left[\rho_{A} \tilde{\mathcal{C}}_{n}^{\dagger} \tilde{\mathcal{C}}_{m}\right].
\end{equation}
Now that we have the truncated correlation matrix, our final task is to derive a relation between the eigenvalues of the reduced density matrix and the eigenvalues of the truncated correlation matrix. We start by substituting (\ref{a13}) in (\ref{a22}) to get:
\begin{equation} \label{a23}
\tilde{C}_{n m}=\operatorname{Tr}\left[K \exp \left(-\mathcal{H}_{A}\right) \tilde{\mathcal{C}}_{n}^{\dagger} \tilde{\mathcal{C}}_{m}\right].
\end{equation}
Substituting (\ref{a15}) in (\ref{a23}) gives:
\begin{equation} \label{a24}
\tilde{C}=K \operatorname{Tr}\left[\exp \left(-2 \sum_{j} \varepsilon_{j} a_{j}^{\dagger} a_{j}+E_{0}\right) \sum_{k l} \Psi_{k}^{+} S_{k n}^{T} S_{m l} \Psi_{l}\right].
\end{equation}
This can be simplified as:
\begin{equation} \label{a25}
\begin{aligned}
\tilde{C} &=K e^{E_{0}} \sum_{k=1}^{2 M} S_{k n}^{T} S_{m k} \operatorname{Tr}\left[\exp \left(-2 \sum_{j} \varepsilon_{j} n_{j}\right) \Psi_{k}^{\dagger} \Psi_{k}\right] \\
&=K e^{E_{0}} \sum_{k \leq M} S_{k n}^{T} S_{m k} \operatorname{Tr}\left[\exp \left(-2 \sum_{j} \varepsilon_{j} n_{j}\right)\left(1-n_{k}\right)\right] 
 + K e^{E_{0}} \sum_{k>M} S_{k n}^{T} S_{m k} \operatorname{Tr}\left[\exp \left(-2 \sum_{j} \varepsilon_{j} n_{j}\right) n_{k}\right],
\end{aligned}
\end{equation}
where $n_{k}=a^{\dagger}_{k}a_{k}$. The value of K can be evaluated by imposing the condition $Tr[\rho_A]=1$, which is a fundamental property of density matrices. We can now uncover the following:
\begin{equation}
\begin{aligned}
\operatorname{Tr}\left[\rho_{A}\right] &=K e^{E_{0}} \operatorname{Tr}\left[\exp \left(-2 \sum_{j} \varepsilon_{j} n_{j}\right)\right]=1 \\
&=K e^{E_{0}}\left(\prod_{l=1}^{M} \sum_{n_{l}=0,1}\right)\left\langle\Phi_{0}\left|\prod_{l=1}^{M} a_{l}^{n_{l}} e^{-2 \varepsilon_{l} n_{l}} a_{l}^{\dagger n_{l}}\right| \Psi_{0}\right\rangle =K e^{E_{0}} \prod_{l=1}^{M} \sum_{n_{l}=0,1} e^{-2 \varepsilon_{l} n_{l}}=K e^{E_{0}} \prod_{l=1}^{M}\left(1+e^{-2 \varepsilon_{l}}\right) \\
&\Longrightarrow K =e^{-E_{0}} \prod_{l=1}^{M} \frac{1}{1+e^{-2 \varepsilon_{l}}}.
\end{aligned}
\end{equation}
Using the above two relations we get:
\begin{equation}
\begin{aligned}
\tilde{C}_{n m} &=\sum_{k \leq M} S_{k n}^{T} S_{m k}\left(1-\frac{e^{-2 \varepsilon_{M+1-k}}}{1+e^{-2 \varepsilon_{M+1-k}}}\right)+\sum_{k>M} S_{k n}^{T} S_{m k} \frac{e^{-2 \varepsilon_{k-M}}}{1+e^{-2 \varepsilon_{k-M}}} \\
&=\sum_{k \leq M} S_{k n}^{T} S_{m k} \frac{1}{1+e^{-2 \varepsilon_{M+1-k}}}+\sum_{k>M} S_{k n}^{T} S_{m k} \frac{1}{1+e^{2 \varepsilon_{k-M}}}.
\end{aligned}
\end{equation}
Comparing this with $H_{nm}$ in (\ref{a14}), which was,
\begin{equation}
\begin{aligned}
H_{n m}&=\sum_{k} S_{n k} \lambda_{k} S_{k m}^{T} 
=\sum_{k \leq M} S_{k n}^{T} S_{m k}\left(-\varepsilon_{M+1-k}\right)+\sum_{k>M} S_{k n}^{T} S_{m k} \varepsilon_{k-M},
\end{aligned}
\end{equation}
where $\lambda_{k}=D_{kk}$. Thus in terms of $\lambda_k$,
\begin{equation}
\begin{aligned}
H_{n m} &=\sum_{k} \lambda_{k} S_{k n}^{T} S_{m k} \\
\tilde{C}_{n m} &=\sum_{k} \frac{1}{1+e^{2 \lambda_{k}}} S_{k n}^{T} S_{m k}=\sum_{k} \xi_{k} S_{k n}^{T} S_{m k}.
\end{aligned}
\end{equation}
We can clearly see that $H$ and $\tilde{C}$ have the same set of eigenvectors and their eigenvalues $\lambda_k$ and $\varepsilon_k$ are related as:
\begin{equation}
\begin{array}{l}
\xi_{k}=\frac{1}{1+e^{2 \lambda_{k}}}, \mspace{50mu}
\lambda_{k}=\frac{1}{2} \log \left(\frac{1-\xi_{k}}{\xi_{k}}\right).
\end{array}
\end{equation}
These relations finally lead us to the equation for entanglement entropy in terms of the eigenvalues of the truncated correlation matrix $\varepsilon_k$:
\begin{equation}
\begin{aligned}
S_{ent} &=-\operatorname{Tr}\left[\rho_{A} \log \rho_{A}\right]=\sum_{k=1}^{M}\left[\frac{2 \varepsilon_{k}}{1+e^{2 \varepsilon_{k}}}+\log \left(1+e^{-2 \varepsilon_{k}}\right)\right] \\
&=-\sum_{k=1}^{M}\left[\xi_{k} \log \xi_{k}+\left(1-\xi_{k}\right) \log \left(1-\xi_{k}\right)\right].
\end{aligned}
\end{equation}
\onecolumngrid
\subsection{ The Keldysh NEGF approach}
\renewcommand{\theequation}{B.\arabic{equation}}
\setcounter{equation}{0}
\subsubsection{Current calculations}
First, we start with the assumption that the leads are large enough so that they can be considered semi-infinite, and thus be effectively modelled by infinite dimensional matrices. This assumption allows using the NEGF approach by partitioning the channel and the leads and working with the channel
Green’s function and incorporating the leads via self energies.
The retarded channel Green’s function can now
be written as:
\begin{equation} \label{b1}
G^{r}(E)=\left[(E+i \eta) I-H_{B d G}-\Sigma_{L}^{r}-\Sigma_{R}^{r}\right]^{-1},
\end{equation}
where E is the free variable energy, and I is the identity
matrix of the dimension of the Hamiltonian, $\eta$ is a small positive damping parameter, and $\Sigma^{r}_L$ and $\Sigma^{r}_R$ represent the retarted self energies for the semi-infinite contacts. 

\begin{figure}[h]
\centering
\includegraphics[width=0.9\textwidth]{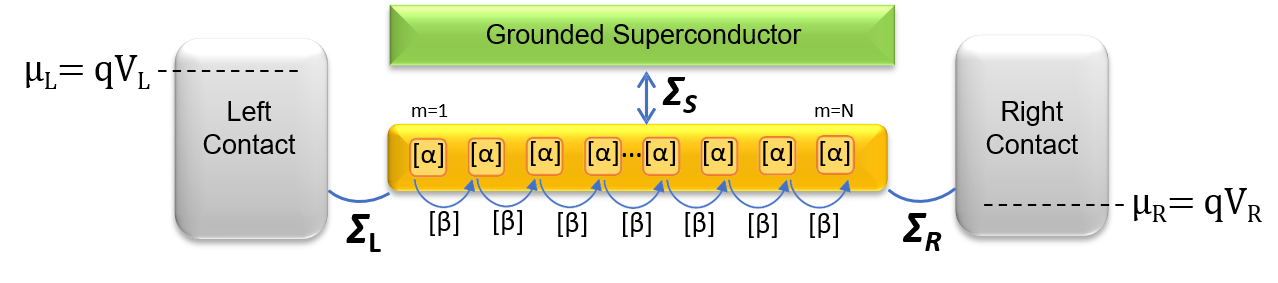}
\caption{NEGF}
\label{fig:8}
\end{figure}

The self energies can be represented in two possible choice of basis: (i) In the eigenbasis of the real-space Hamiltonian, (ii) the eigen basis with wide-band approximation, where the self energy is encapsulated via a parameter $\Gamma_{L/R}$. In the basis of the real-space hamiltonian, the self-energy matrices can be evaluated by a straightforward calculation of the matrix elements $\Sigma^{r}_{\alpha}\left(i,i\right)=\sigma_{\alpha \in L/R}$ :
\begin{equation} \label{eq:b2}
\sigma_{\alpha \in L / R}(E)=\sum_{k} \frac{\left|\tau_{\alpha, k}\right|^{2}}{\left(E-\epsilon_{k \alpha}+i \eta\right)}=P \int d \epsilon_{k \alpha} \frac{\left|\tau_{\alpha, k}\right|^{2}}{\left(E-\epsilon_{k \alpha}\right)}-i \pi \int d \epsilon_{k \alpha}\left|\tau_{\alpha, k}\right|^{2} \delta\left(E-\epsilon_{k \alpha}\right),
\end{equation}
where P stands for the Cauchy principal value. Recasting this integral into a real and an imaginary part ($\sigma_{\alpha}=\epsilon_{\alpha}\left(E\right)-i\gamma_{\alpha}/2$) provides a better physical picture , with the imaginary part $\gamma_{\alpha}=i\left(\sigma^r_{\alpha}-\sigma^a_{\alpha}\right)$ representing the level broadening, and the real part $\epsilon_{\alpha}$  calculated from the principal value integral.

In this work, we stick to the wide-band approximation basis. Under this approximation, $\tau_{k\alpha,m}$ is independent of energy and thus the principal value integral representing the real-part of the self energy vanishes and we are just left with the
broadening matrix $\Gamma_{\alpha}\left(E\right)$ such that the self energies can be represented as $\Sigma^r_{\alpha}= -i\Gamma_{\alpha}/2$, which can be treated as
an input parameter. Once we are done calculating the self energies, we
need to evaluate the “lesser” Green’s function:
\begin{equation} \label{b3}
G^{<}(E)=G^{r}(E)\left(\Sigma_{L}^{<}(E)+\Sigma_{R}^{<}(E)\right) G^{a}(E),
\end{equation}
where $G^a$ is the advanced Green’s function (hermitian conjugate of the retarded Green’s function calculated
from (\ref{b1})). Calculating the lesser Green's function is equivalent to calculating the electron (hole) correlation function. Here, $\Sigma^{<}_{L,R}(E)$ represent the
in-scattering functions from leads L and R respectively which can be evaluated using: 
\begin{equation} \label{b4}
\begin{aligned}
\Sigma_{\alpha}^{<}(E) =-\left[\Sigma_{\alpha}^{r}(E)-\Sigma_{\alpha}^{a}(E)\right] f_{\alpha}(E) 
=i \Gamma_{\alpha}(E) f_{\alpha}(E),
\end{aligned}
\end{equation}
where $f_{\alpha}= f(E-\mu_{\alpha})$ represents the Fermi-Dirac distribution in either leads $\alpha= L(R)$ with a chemical potential $\mu_{\alpha}$. In this formulation, all different components of the currents can then be deduced from the current operator formula. The elements of the current operator matrix can be deduced using the correlation function and the hopping elements of the Hamiltonian. A rigorous but straightforward calculation based on fundamental consideration gives the current operator matrix:
\begin{equation} \label{b5}
\begin{aligned}
I_{L}^{o p}(E)= \frac{e}{h}\left[G^{r}(E) \Sigma_{L}^{<}-\Sigma_{L}^{<} G^{a}(E)\right.
\left.+G^{<}(E) \Sigma_{L}^{a}(E)-\Sigma_{L}^{r}(E) G^{<}(E)\right].
\end{aligned}
\end{equation}
Taking the trace of the above equation gives the net charge current, which is given by the current formula:
\begin{equation} \label{b6}
I_{L}(E)=\frac{i e}{h} \text { Trace }\left[\Gamma_{L}(E) f_{L}(E)\left(G^{r}(E)-G^{a}(E)\right)+\Gamma_{L}(E) G^{<}(E)\right].
\end{equation}
The above formula can be further simplified by the notation
introduced in with the spectral function $A(E) = i(G^r-G^a)$ and the electron correlation matrix $G^n(E) = -iG^{<}(E)$ as
\begin{equation} \label{b7}
I_{L}(E)=\frac{e}{h} \text { Trace }\left[\Gamma_{L}(E) f_{L}(E) A(E)-\Gamma_{L}(E) G^{n}(E)\right].
\end{equation}
Using simple manipulations as given below we get:
\begin{equation} \label{b8}
\begin{aligned}
A(E) &=i\left(G^{r}-G^{a}\right) 
=iG^{r} \left[\left(G^{a}\right)^{-1}-\left(G^{r}\right)^{-1}\right] G^{a} 
=iG^{r} \left[\Sigma^{r}-\Sigma^{a}\right] G^{a} 
=G^{r} \Gamma G^{a},
\end{aligned}
\end{equation}
where $\Gamma(E) = \Gamma_L(E)+\Gamma_R(E)$. Using the above in conjunction with (\ref{b3}) and (\ref{b4}), along with the properties
of the trace operation, yields the Landauer transmission
formula for the current as
\begin{equation} \label{b9}
I_{L}=\int d E T r a c e\left[\Gamma_{L} G^{r} \Gamma_{R} G^{a}\right]\left(f_{L}(E)-f_{R}(E)\right),
\end{equation}
where $f_{L(R)}(E)$ represents the Fermi-Dirac distribution
in either lead. However, in the case involving superconductors, either functioning as a contact or as the central system or both,
the net current through the contact $\alpha$ is the difference
between electron and hole currents in Nambu space. We
can now use the full fledged current operator in (\ref{b5}) to
evaluate the current from first principles as
\begin{equation} \label{b10}
I_{L}^{o p}(E)=\frac{e}{h} \tau_{z}\left[G^{r}(E) \Sigma_{L}^{<}(E)-\Sigma_{L}^{<}(E) G^{a}(E)+G^{<}(E) \Sigma_{L}^{a}(E)-\Sigma_{L}^{r}(E) G^{<}(E)\right],
\end{equation}
where $\tau_z = \sigma_z \otimes I_{N \times N} $ , where  is the Pauli-z matrix and  is the identity matrix of dimension $N\times N$. The next step is to take a trace of the above equation,
which in general need not take the form of (\ref{b6}), typically
when the contacts are superconducting due to the non-diagonal
structure of $\Sigma_{L(R)}$, and may lead to erroneous results, specifically when evaluating the Josephson currents. However, in our case, as the contacts are normal, and hence  diagonal, (\ref{b10}) indeed takes the form of (\ref{b6}) with the net current becoming
\begin{equation} \label{b11}
I^{\alpha}=\int dE \frac{(I_{\alpha}^{(e)}-I_{\alpha}^{(h)})}{2},
\end{equation}
where each current $I^{e(h)}_{\alpha}$ can be evaluated separately using
the form in (\ref{b6}) or (\ref{b7}). The factor of a half comes
since the original Hamiltonian been doubled while writing the BdG Hamiltonian in order to
write it consistently in the Nambu space. It is then instructive
in our context to note that various matrices defined within the approach will have a matrix structure due to the electron-hole Nambu space. In particular, the contact broadening matrices etc., can be written with a general diagonal structure (due to the diagonal structure of normal contacts in Nambu space) as $\Gamma_{\alpha}=\Gamma^{ee}_{\alpha}+\Gamma^{hh}_{\alpha}$, where the superscripts $ee(hh)$ represent the electron (hole) diagonal part of the self energy or broadening matrix. Following this and using the above observations on the current operator
formula in (\ref{b7}), we obtain the electron (hole) current
across the TS as a sum of three components:
\begin{equation} \label{b12}
\begin{aligned}
I_{L}^{e(h)}(E) &=\frac{e}{h}\left(\text { Trace }\left(\Gamma_{L}^{e e(h h)} G^{r} \Gamma_{R}^{e e(h h)} G^{a}\right)\left[f_{L}^{e e(h h)}(E)-f_{R}^{e e(h h)}(E)\right]\right) \longrightarrow (i)\\
&+\frac{e}{h}\left(\text { Trace }\left(\Gamma_{L}^{e e(h h)} G^{r} \Gamma_{L}^{h h(e e)} G^{a}\right)\left[f_{L}^{e e(h h)}(E)-f_{L}^{h h(e e)}(E)\right]\right) \longrightarrow (ii)\\
&+\frac{e}{h}\left(\text { Trace }\left(\Gamma_{L}^{e e(h h)} G^{r} \Gamma_{R}^{h h(e e)} G^{a}\right)\left[f_{L}^{e e(h h)}(E)-f_{R}^{h h(e e)}(E)\right]\right), \longrightarrow (iii)
\end{aligned}
\end{equation}
where the term (i) represents the direct transmission process of
either the electron or the hole, (ii) represents the direct
Andreev transmission and (iii) represents the crossed
Andreev transmission. At this point it is worth noting
that $f^{ee}_{\alpha} = f(E-\mu_{\alpha})$, $f^{hh}_{\alpha} = f(E+\mu_{\alpha})$. The matrix structure of the broadening matrix $\Gamma$ itself is diagonal such that $\Gamma^{ee}_L = \Gamma_L(1,1)=\gamma$, $\Gamma^{hh}_L = \Gamma_L(2,2)=\gamma$, $\Gamma^{ee}_R = \Gamma_R(2N-1,2N-1)=\gamma$, $\Gamma^{hh}_R = \Gamma_R(2N,2N)=\gamma$ and zeros otherwise. The Andreev transmission can then be written as $T_A(E) = \gamma^2 | G^{r,eh}_{11}(E) |^2$ and the crossed Andreeev transmission as $T_{CA}(E) = \gamma^2 | G^{r,eh}_{1N}(E) |^2$
the direct transmission can be written as $T_D(E) = \gamma^2 | G^{r,ee}_{1N}(E) |^2$. Here $G^{r,ee(hh)}_{ij}$
represents the i, j element of the electron (hole) diagonal block of the retarded Green's function in Nambu space and $G^{r,eh(he)}_{ij}$ represents that of the off-diagonal block of the retarded Green's function in Nambu space.

\subsection{Local and non-local conductances}
\subsubsection{Floating superconductor configuration}
We start with the current formula for a generic bias situation, ($V_{L}$,$V_{R}$) for the left and right contacts with $\mu_{L}=eV_{L}$ and $\mu_{R}=eV_{R}$

The current through one of the contacts has contributions from both electron and hole flow. The net current is given by:
\begin{equation}
I^{L}=\frac{I_{L}^{(e)}-I_{L}^{(h)}}{2}.
\end{equation}
The individual electron and hole components can then be derived using:
\begin{equation}
\begin{aligned}
I_{L}^{(e)}=-\frac{e}{h} &\left\{\int d E T_{A}^{(e)}(E)\left[f\left(E-e V_{L}\right)-f\left(E+e V_{L}\right)\right]\right.\\
&+\int d E T_{C A R}^{(e)}(E)\left[f\left(E-e V_{L}\right)-f\left(E+e V_{R}\right)\right] \\
&\left.+\int d w E T_{D}^{(e)}(E)\left[f\left(E-e V_{L}\right)-f\left(E-e V_{R}\right)\right]\right\},
\end{aligned}
\end{equation}

\begin{equation}
\begin{aligned}
I_{L}^{(h)}=-\frac{e}{h} &\left\{\int d E T_{A}^{(h)}(E)\left[f\left(E+e V_{L}\right)-f\left(E-e V_{L}\right)\right]\right.\\
&+\int d E T_{C A R}^{(h)}(E)\left[f\left(E+e V_{L}\right)-f\left(E-e V_{R}\right)\right] \\
&\left.+\int d E T_{D}^{(h)}(E)\left[f\left(E+e V_{L}\right)-f\left(E+e V_{R}\right)\right]\right\}.
\end{aligned}
\end{equation}
Here, $I^{(e)}_L$ and $I^{(h)}_L$ can be shown to be equal and opposite using symmetry conditions and especially:
\begin{equation}
\begin{aligned}
T_{CAR}^{(e)}(E) \equiv T^{(h)}_{C A R}(E), \mspace{50mu}
T_{D}^{(e)}(E) \equiv T_{D}^{(h)}(-E).
\end{aligned}
\end{equation}
It is thus sufficient to consider either one of $I^{(e)}_L$ and $I^{(h)}_L$ while calculating the net current through a contact. Let us now consider the conductance matrix:
\begin{equation}
\mathrm{G}=\left(\begin{array}{cc}
G_{L L} & G_{L R} \\
G_{R L} & G_{R R}
\end{array}\right)=\left(\begin{array}{cc}
\left.\frac{\partial I_{L}}{\partial V_{L}}\right|_{V_{R}=0} & \left.\frac{\partial I_{L}}{\partial V_{R}}\right|_{V_{L}=0} \\
\left.\frac{\partial I_{R}}{\partial V_{L}}\right|_{V_{R}=0} & \left.\frac{\partial I_{R}}{\partial V_{R}}\right|_{V_{L}=0}.
\end{array}\right)
\end{equation}
Now let us consider only the electron current for now since $I_L=I^{(e)}_L$. Using this fact we can evaluate the Local conductance as:
\begin{equation}
\begin{aligned}
G_{L L} &\equiv \left. \frac{\partial I_{L}}{\partial V_{L}} \right|_{V_{R}\equiv 0 }\\
& \equiv \frac { e } { h } \left\{\frac { \partial } { \partial V_L } \left[\int d E T_{A}(E)[f(E-e V_L)-f(E+eV_L)]\right.\right.
+\int d E T_{CAR}(E)[f(E-e V_L)-f(E)] \\
&\mspace{450mu}+\left.\left.\int d E T_{D}(E)[f(E-eV_L)-f(E)]\right]\right\} \\
& \equiv \frac{e}{h}\left[\int d E\right. T_{A}(E)\left[\frac{\partial f\left(E-eV_L\right)}{\partial V_L}-\frac{\partial f(E+e V_L)}{\partial V_L}\right] 
+ \int d E T_{CA R}(E)\left[\frac{\partial f(E-e V_L)}{\partial V_L}\right] \\
&\mspace{450mu} +\left.\int d E T_{D}(E)\left[\frac{\partial f(E-e V_L)}{\partial V_L}\right]\right].
\end{aligned}
\end{equation}
At zero temperature, the $f\left(x\right)=\Theta\left(-x\right)$ where $\Theta$ is the Heaviside step function. This implies $f\left(E-eV \right)=\Theta\left(eV-E\right)$ and similarly we can write the other terms. Since the derivative of the heaviside function is the Dirac Delta function, we get:
\begin{equation}
\begin{aligned} \label{b22}
\left.G_{LL}(V)\right|_{T \rightarrow 0} \equiv \frac{e^2}{h}\left[T_{A}(E=e V)+T_{A}(E=-e V)\right.+
\left.T_{C A R}(E=eV)+T_{D}(E=e V)\right].
\end{aligned}
\end{equation}
Now we can similarly get the non local conductance.
\begin{equation}
\begin{aligned}
G_{LR}=& \left.\frac{\partial I_L}{\partial V_R}\right|_{V_L\equiv 0} \\
=&\frac { e } { h } \left\{\frac { \partial } { \partial V_R } \left[\int d E T_{CAR}(E)[f(E)-f(E+eV_R)]\right.\right.
\left.\left.+\int d  E T_{D}(E)\left[f\left(E\right)-f\left(E-e V_{R}\right)\right]\right]\right\} \\
=& \frac{e}{h}\left[ \int d E T_{CA R}(E)\left[\frac{-\partial f(E+e V_R)}{\partial V_R}\right]\right. 
+\left. \int d E T_{ D}(E)\left[\frac{-\partial f(E-e V_R)}{\partial V_R}\right].
\right]
\end{aligned}
\end{equation}
Once again, using the property of fermi dirac functions as $T\rightarrow 0$ as $f(x)=\Theta(-x)$ we get:
\begin{equation}
\begin{aligned} \label{b24}
\left.G_{LR}(V)\right|_{T \rightarrow 0} \equiv \frac{e^2}{h}\left[T_{D}(E=e V)-T_{CAR}(E=-e V)\right].
\end{aligned}
\end{equation}
\subsubsection{Grounded superconductor configuration}
When the proximitizing superconductor is connected to the ground, current can flow into the superconducting terminal as well. This leads to an additional self-energy term ($\Sigma_S$) in the Green's function. The Green's function for the grounded setup is thus given by :

\begin{equation}
G^{r}(E)=\left[(E+i \eta) I-H_{B d G}-\Sigma_{L}^{r}-\Sigma_{R}^{r}-\Sigma_{S}\right]^{-1},
\end{equation}
where $\Sigma_S$ can be derived using a method similar \eqref{eq:b2} and is given by:
\begin{equation}
\Sigma_S(E)=-\gamma \frac{E+\Delta \tau_{x}}{\sqrt{\Delta^{2}-E^{2}}},
\end{equation}
where $\gamma$ represents the coupling strength between the Kitaev chain and the superconducting contact, and $\tau_x$ represents the Pauli-sigma $x$ matrix.

Due to the presence of the superconducting contact, as we'll see below, an additional term gets added to the previously derived formula for local conductance (\ref{b22}), and no additional term gets added to the non-local conductance formula (\ref{b24}).

Starting with the general current operator for the left contact (\ref{b10}), we get the electron current for the left contact as:
\begin{equation} \label{b27}
\begin{aligned}
I_{L}^{e}(E)&=\frac{e}{h} \operatorname{Tr}\left[\Sigma_{L}^{<}(E)\left(G^{r}(E)-G^{a}(E)\right)\right.
\left.- G^{<}(E)\left(\Sigma_{L}^{a}-\Sigma_{L}^{r}\right)\right] \\
&=\frac{e}{h}  \operatorname{Tr}\left[\Gamma_{L} f_{L}\left(G^{r}(E)\Gamma G^{a}(E)\right)\right.
\left.+ i G^{r}(E)\left(\Sigma_{L}^{<}+\Sigma_{R}^{<}+\Sigma_{S}^{<}\right) G^{a}(E) \Gamma_{L}\right],
\end{aligned}
\end{equation}
where $\Sigma^{<}_S$ is defined by the (\ref{b4}) with the Fermi energy of the superconducting contact given by $f_S$. The broadening matrix $\Gamma$ is the sum of the broadening matrices for left,right and superconducting contacts: $\Gamma = \Gamma_L+\Gamma_R+\Gamma_S$. On further expanding the terms in (\ref{b27}) we get,
\begin{equation}
I_{L}^{e}(E)=\frac{e}{h}\operatorname{Tr}\left[\Gamma_{L} f_{L}G^{r}(E)\left(\Gamma_L+\Gamma_R+\Gamma_S\right)G^{a}(E)+ iG^{r}(E)\left(i\Gamma_{L}f_{L}(E)+i\Gamma_{R}f_{R}(E)+i\Gamma_{S}f_{S}(E)\right)G^{a}(E)\Gamma_{L}\right].
\end{equation}
The additional terms due to the presence of the superconducting contact are:
\begin{equation} \label{b29}
\frac{e}{h}\operatorname{Tr}\left[\underbrace{\Gamma_{L} f_{L}G^{r}(E)\Gamma_SG^{a}(E)}_{(\mathrm{i})}-\underbrace{G^{r}(E)\Gamma_{S}f_{S}(E)G^{a}(E)\Gamma_{L}}_{(\mathrm{ii})} \right].
\end{equation}
The term (i) here has a $V_L$ dependence since $f_L=f(E-eV_L)$, whereas the term (ii) has neither $V_L$ nor $V_R$ dependence. Since the first term has a $V_L$ dependence it adds an additional term to the local conductance formula (\ref{b22}):
\begin{equation}
G^{Gnd}_{LL}(V_L) = \frac{e}{h}\operatorname{Tr}\left[\Gamma_{L} \frac{df_{L}}{dV_L}G^{r}(E)\Gamma_SG^{a}(E)\right].
\end{equation}
Using the fact that the derivative of Fermi function at zero temperature is the Dirac-delta function, we get the additional term to local conductance as:
\begin{equation}
G^{Gnd}_{LL}(V) = \frac{e^2}{h}\operatorname{Tr}\left[\Gamma_{L}G^{r}(V)\Gamma_SG^{a}(V)\right].
\end{equation}
Since neither of the terms in (\ref{b29}) have $V_R$ dependence, no additional term gets added to the non-local conductance equation (\ref{b24}) on introducing the grounded superconductor.

\end{document}